\documentclass[onecolumn,fleqn]{aastex}
\pdfoutput=1
\usepackage{graphicx,subfigure}
\usepackage{bm}
\usepackage{longtable}
\usepackage{amsmath,amssymb,graphicx}
\usepackage{rotating}
\usepackage{color}
\usepackage[normalem]{ulem}
\shorttitle{Reverse shocks and radio afterglows}
\shortauthors{Resmi \& Zhang}
\newcommand{\be}{\begin{equation}}
\newcommand{\ee}{\end{equation}}
\newcommand{\bea}{\begin{eqnarray}}
\newcommand{\eea}{\end{eqnarray}}
\newcommand{\Lb}{\left(}
\newcommand{\Rb}{\right)}

\newcommand{\tp}{t_{\rm{peak}}}
\newcommand{\nuaRS}{{\nu_a}^{\rm RS}}
\newcommand{\numRS}{{\nu_m}^{\rm RS}}
\newcommand{\nuaFS}{{\nu_a}^{\rm FS}}
\newcommand{\numFS}{{\nu_m}^{\rm FS}}
\newcommand{\nucRS}{{\nu_c}^{\rm RS}}
\newcommand{\nucFS}{{\nu_c}^{\rm FS}}
\newcommand{\taumRS}{{\tau_m}^{\rm RS}}
\newcommand{\taumFS}{{\tau_m}^{\rm FS}}
\graphicspath{{./}{Figures/}}
\begin{document}
\title{Gamma Ray Burst reverse shock emission in early radio afterglows}
\author{Lekshmi Resmi \altaffilmark{1}, Bing Zhang \altaffilmark{2}
}
\email{l.resmi@iist.ac.in}
\altaffiltext{1}{Indian Institute of Space Science \& Technology, Trivandrum, India}
\altaffiltext{2}{Department of Physics and Astronomy, University of Nevada, Las Vegas, USA}
%
%





\begin{abstract}
{Reverse shock (RS) emission from Gamma Ray Bursts is an important tool in investigating the nature of the ejecta from the central engine. If the ejecta magnetization is not high enough to suppress the RS, a strong RS emission component, usually peaking in the optical/IR band early on, would give important contribution to early afterglow light curves. In the radio band, synchrotron self-absorption may suppress early RS emission, and also delay the RS peak time. In this paper, we calculate the self-absorbed RS emission in the radio band for different dynamical conditions. In particular, we stress that the RS radio emission is subject to self-absorption in both reverse and forward shocks. We calculate the ratio between the reverse to forward shock flux at the RS peak time for different frequencies, which is a measure of the detectability of the RS emission component. We then constrain the range of physical parameters for a detectable RS, in particular the role of magnetization. We notice that unlike optical RS emission which is enhanced by moderate magnetization, a moderately magnetized ejecta does not necessarily produce a brighter radio RS due to the self-absorption effect. For typical parameters, the RS emission component would not be detectable below 1 GHz unless the medium density is very low (e.g. $n < 10^{-3} ~{\rm cm^{-3}}$ for ISM and $A_* < 5\times 10^{-4}$ for wind). These predictions can be tested with the afterglow observations with current and upcoming radio facilities such as JVLA, LOFAR, FAST, and SKA.}
\end{abstract}

\keywords{gamma ray burst: general, radiation mechanisms: non-thermal}
%
%
\section{Introduction}
Even decades after their discovery, the central engine of Gamma Ray Bursts (GRBs) and the way it powers the explosion still remain largely unsolved \citep{Kumar:2015upa}. Nature of the outflow from the central engine, mainly its composition and degree of magnetization are still not properly understood. Prompt and early multi-wavelength emission is the main messenger to probe the central engine. Early X-ray emission is known to display emission components like flares and ``internal plateaus'' that are usually ascribed to late central engine activities \citep{Zhang:2005fa,Nousek:2005fm,Troja:2007ge}. Early multi-wavelength emission should also contain the emission from the reverse shock that develops once the ejecta encounters the cold medium surrounding the burst \citep{Meszaros:1996sv,Meszaros:1999kv,Sari:1999iz,Sari:1999kj}. A strong reverse shock is developed if the outflow from the central engine is baryonic, i.e. the magnetization parameter $\sigma \leq 0.1$\footnote{When $\sigma$ increases further, a reverse shock becomes weaker gradually due to the enhancement of the magnetic pressure in the outflow, and disappears completely at $\sigma > 1$ when the magnetic pressure in the outflow exceeds the forward shock thermal pressure \citep{Zhang:2004ie,Mizuno:2008qj,Mimica:2008up}.}. Because the ejecta density is higher compared to that of the ambient medium, the reverse shock temperature is lower, leading most of its emission towards frequencies around and lower than the optical band. 

Thanks to the abundant observational data in the optical band, most of the studies of reverse shock emission in the literature so far have focused on the optical/IR band \citep[e.g.][]{Sari:1999kj,Meszaros:1999kv,Kobayashi:2003zk,Kobayashi:2002uw,Zhang:2003wj,Kumar:2003fy,Wu:2003cm,Fan:2004ka,Nakar:2004wf,Zhang:2004ie,Zou:2005sw,Jin:2007uf,Harrison:2012af,Japelj:2014hsa,Gao:2015gqa}, see recent review of \citep{Gao:2015lga}. Due to the improved sensitivity of the JVLA, deep and fast monitoring campaigns of radio afterglows is now possible \citep{Chandra:2011fp,Laskar:2013uza}. Lower frequency studies are possible with LOFAR and in the future with the Five hundred meter Aperture Spherical Telescope (FAST), and the Square Kilometer Array (SKA). In particular, SKA will dramatically improve radio afterglow detection rates or lower the upper limits. A detailed theoretical study of reverse shock radio emission predictions is called for. 

\cite{Sari:1995nm} and \cite{Kobayashi:2000af} have derived the thermodynamical parameters of the reverse shock for a constant density external ambient medium and estimated the synchrotron emissivity. Later, several authors \citep{Kobayashi:2003zk, Wu:2003cm, Kobayashi:2003yh, Zou:2005sw} have extended the formalism to an ambient medium driven by a stellar wind. Recently, \cite{Gao:2013mia} have presented a consolidated review of analytical expressions for reverse shock flux evolution. The difficulty of studying radio reverse shock emission properties lies in correctly accounting for the synchrotron self-absorption effect in many different spectral regimes. Several authors have in the past incorporated self-absorption in the reverse shocked ejecta \citep{Kobayashi:2003zk,Nakar:2004wf,Zou:2005sw}. In particular, \cite{Gao:2013mia} presented a detailed treatment of self-absorption in all possible spectral regimes for both reverse and forward shock emission. In most previous treatments, many authors have nevertheless assumed the reverse shock microphysics to be the same as that of the forward shock for the sake of reducing the number of parameters. However, since the upstream of reverse shock is the ejecta from the central engine, which can be very different from the upstream of the forward shock, i.e. the circumburst medium, it is possible or even likely that the microphysics parameters of the reverse and forward shocks are different. In particular, since most central engine models invoke a strong magnetic field at the engine, the magnetization parameter of the reverse shock can be very different from that of the forward shock. \cite{Zhang:2003wj} found that in order to reproduce bright GRB 990123-like reverse shock emission in the optical band characterized by a $\sim t^{-2}$ decay, the reverse shock should be more magnetized than the forward shock. \cite{Zhang:2004ie} investigated the optical reverse shock emission for an arbitrary magnetization parameter $\sigma$, and found that a moderately magnetized reverse shock (with $\sigma$ slightly below unity) can have magnetic fields strong enough to enhance the reverse shock emission but not strong enough to suppress reverse shock dynamics, which is most favorable for the reverse shock detection in the optical band. Follow-up studies suggest that a moderately magnetized reverse shock is required to interpret a good fraction of GRBs (e.g. \citep{Gomboc:2008ug,Harrison:2012af,Japelj:2014hsa,Gao:2015gqa}). Here we present a comprehensive study of low frequency reverse shock emission in both constant density and wind driven ambient medium profiles with the self-absorption processes fully taken into account. We will examine the variation in light curves depending on the microphysics of the reverse shocked medium, in particular, the consequence of a moderately magnetized reverse shock.

In Section 2 we discuss the basic framework of the external forward-reverse shock model, and introduce the main parameters of the problem. In Section 3, we describe the estimation of the synchrotron self-absorption frequency. Sample lightcurves with varying parameters are presented in Section 4. The conditions for detectable reverse shock emission are presented in Section 5. The results are summarized in Section 6 with discussion.
\section{Reverse Shock Emission}
Gamma Ray Bursts are explosions powered by a central engine which is believed to launch a collimated outflow of ultra-relativistic ejecta. When the ejecta encounters the circumburst medium, a two-shock system gets developed. A forward shock (FS) moves into the  circumburst medium and a reverse shock (RS) propagates back to the ejecta. The reverse shock starts off as Newtonian and becomes relativistic if the ejecta is thick enough. In the standard framework, the RS dynamics is considered in two asymptotic regimes: the Newtonian (thin shell) and the relativistic (thick shell) \citep{Sari:1995nm} (SP95), \citep{Kobayashi:2000af} (K00). A critical condition is whether the blast wave accumulates a mass of $M_{ej}/\eta$ (where $M_{ej}$ is the total ejected mass in the shell and $\eta$ is the initial lorentz factor of the ejecta) before the reverse shock becomes relativistic. 
This can be translated to a comparison between the time ($t_\gamma$) it takes for the FS to accumulate $M_{ej}/\eta$ mass and the duration of the burst itself ($T$), i.e. $t_\gamma > T$ for a non-relativistic RS (or thin shell) and $T> t_\gamma$ for a relativistic RS (or thick shell). The shock crossing time can be in general written as $t_\times = \mbox{max}(t_\gamma,T)$. See SP95 and K00 for a detailed discussion of these timescales.

%
Once the reverse-forward shock system is developed, four distinct regions can be defined. Region-1 is the unshocked ambient medium, Region-2 is the forward shocked shell, Region-3 is the reverse shocked ejecta, and Region-4 is the unshocked ejected shell. The fundamental parameters characterizing the RS dynamics are (i) the ratio $f(r)$ between the ejected shell density $n_4(r)$ and the ambient medium density $n_1(r)$,
\be
f(r) = n_4(r)/n_1(r),
\label{eq01}
\ee 
 (ii) the initial bulk Lorentz factor $\eta$ and (iii) thickness $\Delta_0$ of the ejected shell, which is related to the observed burst duration $T$ as $\Delta_0 = c T/(1+z)$ where $z$ is the redshift of the burst. 

However, $f(r)$ can be written in terms of the total kinetic energy $E$, external medium density $n_1(r)$, and $\Delta_0$, parameters that are directly related to prompt and forward shock emission. For this purpose we first rewrite $n_4(r)$ in terms of the ejected mass $M_{\rm ej}$. $M_{\rm ej} = 4 \pi r^2 {\tilde{\Delta}} n_4(r)$, where ${\tilde{\Delta}}$ is the shell thickness in the co-moving frame of the unshocked shell (and is related to the observer frame thickness $\Delta_0$ as ${\tilde{\Delta}} =  \Delta_0 \eta$). For a thick shell one assumes that the shell thickness remains constant throughout, or in other words, the shell does not spread (K00). In the case of thin shell, the shell spreads as $\Delta(r) = r/\eta^2$. 

To rewrite $M_{\rm ej}$ in terms of $E$ and $n_1(r)$, it is convenient to introduce the Sedov length ($l$). Sedov length is the radius of the fireball when the rest energy $m(l) c^2$ of the material that is swept up equals the total energy $E$ of the ejecta. Total swept up mass $m(l)$ at $l$ can be written as 
\be
m(l) = \left\{ \begin{array}{rl}
(4 \pi/3) l^3 n_0 m_p & \mbox{ISM,} \\
4 \pi l r_0^2 n_0 m_p & \mbox{wind,}
\end{array} \right.
\label{eq2}
\ee
where $n_0$ is used to rewrite $n_1(r)$. For a constant density ambient medium, $n_0 = n_1/(1 \mbox{atom}/\mbox{cc})$. The wind driven ambient medium is assumed to be of a density structure $n_1(r) = n_0 (r/r_0)^{-2}$. In terms of the usual parametrization of wind density profile by \cite{Chevalier:1999jy}, $n_0 = 5 \times 10^{11} (A_{\star}/m_p ) r_0^{-2}$, where $A_\star$ of unity corresponds to the density of the medium formed due to a constant mass loss rate of $10^{-5}$~$\mbox{M}_{\sun}/\mbox{yr}$, driven by a wind of $1000$~km$\rm{s}^{-1}$ velocity. From Eq-\ref{eq2}, one has
\be
l = \left\{\begin{array}{rl}
(3 E /(4 \pi n_0 m_p c^2))^{1/3} & \mbox{ISM,} \\
E/(4 \pi r_0^2 n_0 m_p c^2) & \mbox{wind.}
\end{array} \right.
\ee
We now use the fact that $M_{ej}$ is also equal to $E/\eta c^2$, i.e., $m(l)/\eta$, and obtain $f$ as a function of $r$ by equating $M_{ej}$ with $m(l)/\eta$. 
\be
f(r) = \left\{ \begin{array}{rl}
 (l/r)^3  & \mbox{thin ISM,} \\
 l^3/\Delta_0 \eta^3 r^2  &\mbox{ thick ISM,} \\
 l/r  & \mbox{ thin wind,} \\
 l/\Delta_0 \eta^2  &\mbox{ thick wind.} 
       \end{array} \right.
\ee
Here $f(r)$ is defined before the shock crossing time ($t_\times$) (i.e., till the unshocked ejecta diminishes). It is important to note that for thin shell cases, $f$ does not depend on $\Delta_0$ after the shell starts to spread. In other words, in the basic equations of Newtonian reverse shock, the initial shell width $\Delta_0$ does not enter. For a relativistic RS (thick shell) in a wind driven ambient medium, $f(r)$ is a constant as both $n_4$ and $n_1$ depend on $r$ in the same way.

We use $f(t_\times) = f(r=r_\times)$, for normalization: i.e. $f(t_\times) = \eta^2$ for thin shell for both types of ambient medium; $f(t_\times) = \eta^{-2} (l/\Delta_0)^{3/2}$ for thick shell ISM, and $f(t_\times) = l/(\Delta_0 \eta^2)$ for thick shell wind.

In order to estimate the synchrotron emission from the reverse shocked region, we need the number density ($n_3$), energy density ($e_3$), total number of radiation electrons ($N_e$) and bulk Lorentz factor $\Gamma_{31}$ of the shocked region. All these can be obtained once $f(r)$ is known. Next, we derive these quantities in two phases: before and after shock crossing.

\subsection{Before shock crossing ($t \le t_\times$)}
The jump conditions at the forward and the reverse shock place a constraint on $f(r)$ in terms of the Lorentz factors of the shocked medium (see SP95 for detailed expressions of the shock jump conditions). Assuming pressure equilibrium at the contact discontinuity between the shocks which is valid as long as the shock crossing is fast\footnote{This assumption becomes a bad approximation if the reverse shock is long-lived \citep{Beloborodov:2006qt,Uhm:2010js}.}, we get,
\be
f =\frac{ 4 \Gamma_{21}^2 (\hat{\gamma}-1) }{(\hat{\gamma} \gamma_{34} +1) (\gamma_{34}-1) },
\label{eq5}
\ee
where $\Gamma_{21}$, the bulk lorentz factor of the shocked ambient medium, is assumed to be much greater than unity; $\gamma_{34}$ is the bulk Lorentz factor of the reverse shocked ejecta with respect to the unshocked ejecta; $\hat{\gamma}$ is the ratio of specific heats of the shocked downstream and varies from $5/3$ for non-relativistic temperatures to $4/3$ for relativistic temperatures. The general expression for $\hat{\gamma}$ is given as (\cite{Uhm:2010js} and references therein),
\be
\hat{\gamma} = \frac{4 \gamma_{34} + 1}{3 \gamma_{34}}.
\ee
While writing Eq-\ref{eq5}, we have assumed that the bulk Lorentz factor of shocked ejecta ($\Gamma_{31}$) is equal to $\Gamma_{21}$. 
$\Gamma_{21}$ is related to $\eta$ and $\gamma_{34}$ by 
\be
\Gamma_{21} = \eta( \gamma_{34} - \sqrt{\gamma_{34}^2 - 1}).
\label{eq7}
\ee
This expression has been derived using Lorentz transformation of velocities and by assuming $\eta \gg 1$. 

From Eq-\ref{eq5} and Eq-\ref{eq7}, $\gamma_{34}$ and $\Gamma_{21}$ can be derived numerically in terms of $f$ and $\eta$ \citep{Nakar:2004wf}. However, in our calculations we use the approximate solutions for $\gamma_{34}$ and $\Gamma_{21}$ derived by SP95. For the thick shell case, one has,
\be
\gamma_{34} = \sqrt{\frac{\eta}{2}} \, \frac{1}{f^{1/4}},
\ee
and
\be
\Gamma_{21} = \Gamma_{31} = \sqrt{\frac{\eta}{2}} \, f^{1/4}.
\ee
For thin shell, one has,
\be
\gamma_{34} = 1+ \frac{\eta^2}{f},
\ee
and
\be
\Gamma_{21} = \Gamma_{31} = \eta \Lb1-\sqrt{\frac{\eta^2}{2 f}}\Rb.
\ee
We can now obtain the number density $n_3 = 4 \gamma_{34} n_4$ and the thermal energy density $e_3 = (\gamma_{34}-1) n_3 m_p c^2$ of the shocked ejecta in terms of $\eta$, $f(r)$ and $\Gamma_{31}$. 

For the total number of shocked electrons, the only remaining parameter required to estimate the synchrotron flux, one has (K00)
\be
N_e = N_\times \frac{t}{t_x},
\ee
for thick shell (both wind and ISM),
\be
N_e = N_\times \Lb \frac{t}{t_x} \Rb^{1/2},
\ee
for thin shell wind, and
\be
N_e = N_\times \Lb \frac{t}{t_x} \Rb^{3/2},
\ee
for thin shell ISM. In the above equations, $N_\times = E/(\eta m_p c^2)$ is the number of shocked electrons at $t_\times$ and $t = r/(2 {\Gamma_{31}}^2 c)$ is the observer time.

\subsection{After shock crossing ($t > t_\times$)}

Analytical treatments of the evolution of the RS medium after shock-crossing is heavily approximated. One assumption that is often used, which we follow here, is that after shock crossing the reverse shocked ejecta achieves the Blandford-McKee (BM) profile \citep{Kobayashi:1999ei,Kobayashi:2000af}. For the relativistic reverse shock it leads to $\Gamma_{31} \propto r^{k-7/2}$, where $k$ is the radial profile index of the ambient medium ($k=2$ for wind and $k=0$ for ISM).
 \cite{Kobayashi:1999ei} notices that for the non-relativistic RS, $\Gamma_{31}$ does not need to follow the BM profile. Instead it follows $\Gamma_{31} \propto r^{-g}$, where $g$ takes a range of values, e.g. $3/2 < g < 7/2$ for ISM \citep{Kobayashi:1999ei}. Nevertheless, as mentioned in \cite{Kobayashi:1999ei}, this range in $g$ does not give rise to a large range in the temporal index of the flux.

The remaining assumption is about the thermodynamics of the reverse shocked ejecta. For a relativistic RS we assume that the sound speed $c_s \sim c/\sqrt{3}$, since protons become relativistic. Since the co-moving width ($\Delta^{\prime}$) of the shock increases as $c_s r/(\Gamma_{31} c)$, the number density scales as $n_3 = N_e/(4 \pi r^2 \Delta^{\prime}) \propto \Gamma_{31}/r^3$. For a non-relativistic RS, protons need not be relativistic, hence we consider $c_s \sim \sqrt{p_3/\rho_3}$, where $p_3$ is the pressure and $\rho_3$ the mass density of the reverse shocked ejecta. Assuming an adiabatic equation of state, $p_3 \propto n_3^{\hat{\gamma}}$, we can obtain both $n_3$ and $p_3$. Once $p_3$ is known, $e_3$ can be obtained as $e_3 = p_3 (\hat{\gamma}-1)$. Using pre shock-crossing relations (Section 2.1) to normalize $n_3$ and $e_3$ at $t_\times$, we obtain, 
\begin{equation}
\begin{array}{rl}
n_3 & = n_3(t_\times) \Lb \frac{r}{r_\times} \Rb^{-6(3+g)/7} \\ 
e_3 & = e_3(t_\times) \Lb \frac{r}{r_\times} \Rb^{-8(3+g)/7},
\end{array}
\end{equation}
for thin shell. We have used $g=2$ for ISM and $g=1$ for wind, and $r_\times = r(t=t_\times)$. For the thick shell, one has
\be
\begin{array}{rl}
n_3 & = n_3(t_\times) \Lb \frac{r}{r_\times} \Rb^{3+k-7/2} \\ 
e_3 & = e_3(t_\times) \Lb \frac{n_3}{n_3 (t_\times)} \Rb^{4/3},
\end{array}
\ee
Number of shocked electrons $N_e$ remains the same as $N_\times$ for $t > t_\times$ in all cases.  
We refer to K00 for a detailed description of the uncertainties involved in the dynamics and thermodynamics of Region-3 post shock-crossing.

\subsection{Microphysics of reverse shock and synchrotron spectrum}
Once we know $n_3$, $e_3$, $N_e$ and $\Gamma_{31}$, we can calculate synchrotron emission from the RS at any given time provided we also know the microphysics of the shocked region that decides the energy content in magnetic fields and electrons. We follow the standard approach in which magnetic fields and electrons in the shocked medium are assumed to carry a constant fraction of the post-shock thermal energy density. The co-moving magnetic field strength in the shocked medium is $B = \sqrt{8 \pi \epsilon_{\rm B,RS} e_3}$, where $\epsilon_{\rm B,RS}$ is the fractional energy content in the magnetic field. Energy density $u_e$ in non-thermal electrons is, $u_e = \epsilon_e e_3$. In order to quantify the differences of microphysics parameters in the two shocked regions, we define the ratios\footnote{These ratios were first defined by \cite{Zhang:2003wj}. Our definition of $R_{\rm B}$, however, is different from \cite{Zhang:2003wj}, who defined $R_{\rm B}$ as the ratio between the RS and FS magnetic field strength. Therefore $R_{\rm B}$ defined in this paper is the square of $R_{\rm B}$ defined in \cite{Zhang:2003wj}.}
\be
\begin{array}{rl}
\mathcal{R}_e = \frac{\epsilon_{\rm e,RS}}{\epsilon_{\rm e,FS}},\\  
\mathcal{R}_{\rm B} = \frac{\epsilon_{\rm B,RS}}{\epsilon_{\rm B,FS}}.
\end{array}
\ee

Assuming the electrons are distributed as a power-law in energy with index $p$, which we take to be the same as that of the forward shock, the minimum Lorentz factor of the electron distribution can be derived as $\gamma_m =\epsilon_{\rm e,RS} \frac{p-2}{p-1} \frac{e_3}{n_3 m_e c^2}$. 

The cooling break $\gamma_c$ in the electron spectrum before shock crossing is $\frac{6 \pi m_e c}{\sigma_T} \frac{1}{2 B^2 \Gamma_{31} t}$ \citep{Sari:1997qe}. After shock crossing, $\gamma_c$ becomes a cut-off Lorentz factor since no new electrons are added further, and follows the same temporal evolution as $\gamma_m$ \citep{Wu:2003cm}. 

The synchrotron spectral parameters $\nu_m$ and $\nu_c$ are derived from $\gamma_m$ and $\gamma_c$, respectively, using the expression $\nu_{m,c} (\gamma_e) = \frac{e}{2 \pi m_e c} B \gamma_{m,c}^2 \Gamma_{31}$. The peak flux $f_m$ (at $\nu_m$ for slow cooling and at $\nu_c$ for fast cooling) is $f_m = \frac{\sqrt{3} e^3}{m_e c^2} \, B\, \frac{N_e \Gamma_{31}}{4 \pi d_L^2}$ \citep{Wijers:1998st}. For forward shock, the same expressions hold for $\nu_{m,c}$ and $f_m$ with $\Gamma_{31}$ replaced by $\Gamma_{21}$. The synchrotron spectrum $f_{\nu}$ is assumed to be a combination of broken power-laws with index $1/3$ for $\nu < \nu_m$, $-(p-1)/2$ for $\nu_m < \nu < \nu_c$, $-1/2$ for $\nu_c < \nu < \nu_m$ and $-p/2$ for ${\rm max}(\nu_m, \nu_c) < \nu$. In the numerical calculations, we introduce a smoothing to the spectral breaks.

\section{Estimation of synchrotron self-absorption frequency}
The synchrotron spectrum described in the previous section is optically thin. However, due to self-absorption, the spectrum is expected to be modified in the low frequency regime. We derive the self-absorption frequency using two different methods: the optical depth method and the blackbody method (see below). Comparing the extent of absorption in RS and FS medium, we see that the RS self-absorption frequency is much higher than that of the FS.

\subsection{Optical depth method}
The optical depth $\tau_{\nu} = \alpha'_{\nu'} \Delta^{\prime}$ and self absorption frequency $\nu_a$ corresponds to $\tau_\nu = 1$. We obtain $\nu_a$ by solving for $\tau_\nu$ using $\alpha'_{\nu'}$ from \cite{ryb} and $\Delta^{\prime} = N_e/(4 \pi n_3 r^2)$.

We normalize the frequency variation of $\tau_{\nu}$ in terms of $\tau_{\nu_p}$ ($\tau_{\nu_m}$ for slow cooling and $\tau_{\nu_c}$ for fast cooling), the optical depth at $\nu_p$ ($\nu_m$ for slow cooling and $\nu_c$ for fast cooling). This gives
\be
\tau_{\nu}/\tau_{\nu_p} = \left\{ \begin{array}{ll}
(\nu/\nu_p)^{-5/3}, & \, \, \nu < \nu_p, \\ \nonumber
(\nu/\nu_p)^{-(p+4)/2}, & \, \, \nu_p < \nu < \nu_\star, \\ \nonumber
(\nu_\star/\nu_p)^{-(p+4)/2} \, (\nu/\nu_\star)^{-(p+5)/2}, & \, \, \nu_\star < \nu, \nonumber
\end{array} \right.
\ee
where $\nu_\star = \mbox{max}(\nu_m,\nu_c)$. 

Hence $\nu_a$ can be expressed as,
\be
\nu_a = \left\{ \begin{array}{ll}
\nu_p \, {\tau_{\nu_p}}^{3/5}, & \, \, \nu_a < \nu_p  \\ 
\nu_p \, {\tau_{\nu_p}}^{\frac{2}{p+4}},  & \, \, \nu_p < \nu_a < \nu_\star  \\
\nu_p \, {\tau_{\nu_p}}^{\frac{2}{p+5}} \, {(\frac{\nu_p}{\nu_\star})}^{\frac{p+4}{p+5}},  & \, \, \nu_\star < \nu_a.
\end{array} \right.
\label{nuaeq}
\ee
\subsubsection{Comparison of self-absorption between FS and RS}
For the reverse shocked medium, $\tau_{\nu_p}$ can be written as,
\be
\tau_{\nu_p} {\rm (RS)} = \frac{\sqrt{3}}{8}\,  3^{p/2} \Gamma{[(3p+2)/12]} \Gamma{[(3p+22)/12]} e \frac{N_e}{r^2} (p-1) \gamma_p^{-5} B^{-1}.
\ee
where $\Gamma[Z]$ is the Euler Gamma function.
Similarly, the optical depth estimated for the forward shock gives
\be
\tau_{\nu_p} {\rm (FS)} = 2 \sqrt{3} \pi 3^{p/2} \Gamma{[(3p+2)/12]} \Gamma{[(3p+22)/12]} e n_1(r) \frac{r}{a} (p-1) \gamma_p^{-5} B^{-1}.
\ee
Here we have assumed the co-moving frame thickness of the downstream of the forward shock to be $r/(a \Gamma_{21})$, where $a$ is a numerical factor of the order of $10$. The number density of the upstream medium, $n_1(r)$, is rewritten for wind and ISM as explained in section-2.

In order to compare the extent of self-absorption in the forward and reverse shock spectrum, we introduce $\mathcal{R}_{\tau}$, the ratio of optical depths of the RS and FS at their corresponding peak frequency $\nu_p(t_\times)$ at the shock crossing time $t_\times$, i.e.,
\be
\mathcal{R}_{\tau} = \frac{\tau_{\rm RS}(t_\times)}{\tau_{\rm FS}(t_\times)} = \frac{1}{4} a \frac{N_e}{4 \pi {r_\times}^3 n_0(r_\times)} \Lb \frac{\gamma_m^{\rm RS}}{\gamma_m^{\rm FS}} \Rb^{-5} \Lb \frac{B^{\rm RS}}{B^{\rm FS}} \Rb^{-1},
\ee
where $r_\times$ is the radius of the fireball at shock crossing. The $\gamma_m$-ratio and $B$-ratio between RS and FS depend on $\mathcal{R}_e$ and $\mathcal{R}_B$, respectively, as well as $e_3/e_2$ and $n_3/n_2$. Assuming pressure equilibrium implies $e_3/e_2 = 1$. In the following, we use approximate expressions for $n_3/n_2$ to get a rough estimate of $\mathcal{R}_\tau$. These approximations are not used in the code for calculating the lightcurves. 

For a thin shell, one has $n_3/n_2 \sim \eta$ and $M_{ej} = \eta m(t_\times) \sim \eta 4 \pi {r_\times}^3 n_0(r_\times) \sim 3 N_e$. Hence $\mathcal{R}_{\tau}$ for the thin shell for both ISM and wind profiles is,
\be
\mathcal{R}_{\tau} \sim \eta^6  \, {{\mathcal R}_e}^{-5} {{\mathcal R}_{\rm B}}^{-1/2}.
\ee

For a thick shell, one has $N_e/(4 \pi n_0(r_\times) {r_\times}^3) \sim  \Gamma_{31}^2 (t_\times)/\eta$ for both types of ambient media. Since the shock is relativistic, one has $n_3(t_\times)/n_2(t_\times) = \gamma_{34} (t_\times) f(t_\times)/\Gamma_{21}(t_\times)$. After employing the relations given in Section-2 for the Lorentz factors and $f$ at $t_\times$, this reduces to ${\Gamma_{31}}^2 (t_\times)/\eta$ for both types of ambient media except minor differences in the numerical factors. Hence, for thick shell,  we can write $\mathcal{R}_{\tau}$ as,
\be
\mathcal{R}_{\tau} = (\Gamma_{31}^2 (t_\times)/\eta)^6 \, {{\mathcal R}_e}^{-5} {{\mathcal R}_{\rm B}}^{-1/2}.
\ee

From Eq-9 it follows that $\Gamma_{31}(t_\times) < \sqrt{\eta}$, consequently, $\mathcal{R}_{\tau}$ is lower for thick shells than for thin shells for the same values of $\eta$, ${{\mathcal R}_e}$ and ${{\mathcal R}_{\rm B}}$. However, for both thin and thick shells, the denser RS materials are much more optically thick than the FS materials, as expected. 

\subsection{Blackbody method}

In another widely used method, self-absorption frequency is determined by equating the blackbody flux with the synchrotron flux \citep{Sari:1999kj,Kobayashi:2003zk}. More specifically, here the synchrotron surface flux (${F_a}^{\prime}$) in the co-moving frame at $\nu_a^{\prime}$ is equated to the flux of a blackbody at temperature $T$ in the Rayleigh-Jeans regime. The temperature $T$ is assumed to be max$(\gamma_a, \gamma_m) \times m_e c^2/k_{\rm B}$, where $\gamma_a$ is the Lorentz factor of an electron radiating with typical synchrotron frequency $\nu_a^{\prime}$.

For illustration, we only consider the case $\nu_a < \nu_m < \nu_c$. Using $F_a^{\prime} = F_m^{\prime} (\nu_a^{\prime}/\nu_m^{\prime})^{1/3}$ and $4 \pi d_L^2 f_m \nu_m = 4 \pi r^2 \Gamma_{31}^2 F_m^{\prime} \nu_m^{\prime}$, we can finally write
\be
8 \pi \gamma_m m_e c^2 (\nu_a^{\prime}/c)^2 = f_m \, (d_L/r)^2 \, \frac{1}{\Gamma_{31}} \, (\nu_a^{\prime}/\nu_m^{\prime})^{1/3}.
\ee
The above equation can be easily solved for $\nu_a$ for the various RS dynamics and ambient medium we considered earlier. The $\nu_a$ thus obtained differs from the $\nu_a$ derived using the optical depth method by a factor $3^{0.3 p} \, (p-1)^{3/5} \Gamma[(3p+2)/12] \Gamma[(3p+22)/12]$. For $p \sim 2.2 - 2.5$, the $\nu_a$ derived using this method is about one-third the $\nu_a$ derived from the optical-depth method. The difference arises because some $p$ dependent terms are not considered in the blackbody method. This result is in agreement with the findings of \cite{Shen:2009mt}, who have investigated self-absorption in detail for prompt emission. 

In Appendix \ref{appen1}, we present the expressions of the RS self-absorption frequency for various spectral regimes and density profiles, both before and after shock crossing.

In the optically thick regime, the synchrotron spectrum is modified as $\nu^2$ for $\nu_a < (\nu_m,\nu_c)$ and $\nu^{5/2}$ for $\nu_a > (\nu_m,\nu_c)$.

\subsection{Pile-up of electrons due to strong synchrotron cooling}

In certain parameter regimes, the reverse shock emission in a wind medium can encounter the condition of strong synchrotron cooling ($\nu_c < \nu_a$). Under such a condition, electrons would pile up at a certain energy due to the balance of synchrotron cooling and self-absorption heating, so that the spectrum is modified to have a quasi-thermal component in the low energy regime \citep{Kobayashi:2003yh}. To obtain an exact description of the electron energy spectrum, full numerical calculations are required \citep{Ghisellini:1996fq}. \cite{Gao:2012sq} used an analytical approximation of this effect within the context of synchrotron and SSC spectrum of GRB prompt and afterglow emission. In our spectrum and lightcurve calculations, we adopt this method in the strong cooling regime.

\section{Radio Lightcurves}
In this section we compute the radio afterglow lightcurves for the various cases we have considered so far.

Figures 1-8 are radio light curves in four frequencies: $22$~GHz, $5$~GHz, $150$~MHz, and $50$~MHz, denoted by $\nu_{\rm obs}$. We consider the ISM (Fig.1-4) and wind (Fig.5-8) cases, for both thin and thick shells. In order to explore a redshift dependence of the results, for each set of parameters, we calculate the light curves for two characteristic redshifts: $z=1$ and $z=5$. We obtain the lightcurves for the four bands for various values of $\mathcal{R}_e$ and $\mathcal{R}_B$. Other parameters ($E_{52}$, $n_0$ or $A_\star$, $\eta$, $\epsilon_e$, $\epsilon_B$, $\theta_j$, $p$,$T_{90}$) are fixed while ensuring that the thin shell and thick shell conditions are satisfied. For comparison, the FS light curves are plotted as the dashed curve. 

A higher $\mathcal{R}_{\rm B}$ leads to higher values for ${f_m}^{\rm RS}, \nuaRS$, and $\numRS$, and a much lower value for $\nucRS$. Increasing $\mathcal{R}_e$ leads to an increase in $\numRS$ and $\nuaRS$. RS lightcurves in GHz range frequencies peak when the fireball becomes optically thin ($\nu_{\rm obs} =\nuaRS$). Increasing $\nuaRS$ delays this peak. Hence, as $\mathcal{R}_e$ and $\mathcal{R}_{\rm B}$ increases, the peak shifts to a later time. Since the RS lightcurve is in the rising phase before the peak, a delayed peak can potentially lead to a higher RS flux. However, if the FS flux is also rising, a delayed peak need not necessarily lead to a higher dominance of RS over FS, as seen in the ISM lightcurves. For the wind lightcurves, where FS flux is nearly constant over time (due to the specific spectral regime the FS is in), a delayed peak results in a mild dominance of RS over FS. This is in contrast to the optical ranges where a higher $\mathcal{R}_B$ always leads to a high RS flux \citep{Zhang:2003wj,Zhang:2004ie}.

In lower radio frequencies (MHz range), the fireball remains optically thick during the peak, and the peak is due to a transition in the effective optical depth. During the fireball evolution, the RS medium eventually becomes optically thin but the FS medium is still optically thick to low frequency RS photons, leading to a change in the effective optical depth. That is, $\nuaRS = \nuaFS$. This transition results in a change in the slope of MHz range lightcurves from a rising to a falling one, thereby resulting in a peak. This peak is always at a later epoch, often an order of magnitude later than the GHz peaks. By this time, ${f_m}^{\rm RS}$ would be going down steeply, as opposed to ${f_m}^{\rm FS}$, which is either constant or varying at a much slower rate. Self-absorption of the flux and the reduction in ${f_m}^{\rm RS}$ together make RS always negligible compared to FS at lower frequencies. 

An achromatic break is seen in the lightcurves due to the jet edge becoming visible (geometric jet break). The break is noticeable in the falling part of the GHz and in the rising part of the MHz lightcurves. Regarding the redshift dependency, the higher redshift lightcurve is fainter, as expected.

\section{Reverse shock detectability from RS and FS flux comparison}
The reverse shock can be detected if it rises above the forward shock, which is most likely to happen around the RS peak time. A measure of RS detectability is the ratio $\chi$ between RS and FS flux at the peak time ($t_{\rm peak}$) of the RS flux. RS detectability implies
\be
\chi = F^{\rm{RS}}({\nu_{\rm{\rm obs}}},\tp)/F^{\rm{FS}}({\nu_{\rm obs}},\tp) > 1.
\label{chieq}
\ee

The parameter $\chi$ depends on the RS and FS spectral regimes of the observing frequency $\nu_{\rm obs}$ at $\tp$. We now derive some analytical expressions for $\chi$ for $\nu_{\rm obs}$ in GHz and MHz ranges corresponding to specific combinations of RS and FS spectral regimes. For simplicity, we only consider the more probable slow cooling case in the analytical calculations. We have also done numerical calculations which can take care of different spectral regimes including fast cooling cases.

\subsection{High radio frequencies}
Following Eq-\ref{nuaeq}, the self absorption frequency can be written in terms of the optical depth $\tau_m$ at $\nu_m$,
\be
\nu_a = \nu_m {\tau_m}^{1/\mu},
\label{eqnua}
\ee
where the index $\mu$ is given by
\be
\mu = \left\{ \begin{array}{ll}
 (p+4)/2, & \nu_a > \nu_m, \\ \nonumber
 5/3, & \nu_a < \nu_m.
 \end{array} \right.
\ee
In some ranges of the parameters, especially when $\mathcal{R}_{\rm B}$ is high, self-absorption frequency can be greater than cooling frequency. However, this usually happens well past $\tp$. Therefore we have not considered this case in the analytical calculations. Nevertheless, in the code and in Fig. 10-13 this is taken care of (see section-3.4).

The high radio frequency (GHz range) lightcurve peaks when the RS medium becomes optically thin to the observed frequency ($\nu_a^{\rm RS} = \nu_{\rm obs}$, or $\tp = t_a$). For typical ranges of physical parameters, the RS spectral breaks at $t_a$ are ordered as $\numRS < \nuaRS < \nucRS$. Using Eq-\ref{eqnua}, the condition $\numRS < \nuaRS$ can be rewritten as
\be
1 < \left( \taumRS \right)^{1/\mu}. 
\ee

To calculate $\chi$, next we consider the FS spectral regime. For standard parameters, the FS spectral breaks at $t_a$ are in the sequence $\nuaFS < \numFS < \nucFS$, with $\numFS$ in the THz ranges. That is, $\nu_{\rm obs} < \numFS$. Since $\nu_{\rm obs} = \nuaRS$ at $t_a$, we can rewrite this as $\nuaRS < \numFS$. Substituting Eq-\ref{eqnua} for $\nuaRS$, one gets
\be
\left( \taumRS \right)^{1/\mu} < \numFS/\numRS.
\ee

Combining the conditions on both FS and RS spectral regimes,
\be
1 < \left( \taumRS \right)^{1/\mu} < \numFS/\numRS.
\label{eq36}
\ee

A consequence of the assumptions on RS and FS spectral regimes is $\nu_a^{FS} < \nu_a^{RS}$, i.e., the reverse shock optical depth is higher than the forward shock optical depth. We treat the opposite condition in the low frequency analysis (next section).

A sketch of the RS and FS spectra at $t_a$ are given in Fig. 9 (left panel).

Finally, the RS and FS fluxes at $\tp$ can be obtained as,

\be
F^{\rm{RS}}({\nu_{\rm obs}},t_a) = {F_m}^{RS}\Lb \frac{\nuaRS}{\numRS} \Rb^{-\beta},
\ee
where $\beta = (p-1)/2$, and

\be
F^{\rm{FS}}({\nu_{\rm obs}},t_a) = {F_m}^{FS} \Lb \frac{\nu_m^{RS}}{\nu_m^{FS}} \Rb^{1/3} \tau_m^{1/3\mu}.
\ee


\subsection{Low radio frequencies}
For the MHz range radio frequencies, which are relevant to SKA and LOFAR, the RS peak is when the RS optical depth falls below that of FS, and the FS medium starts to become responsible for the absorption of RS photons ($\nu_a^{\rm RS} = \nu_a^{\rm FS}$ or $\tp = t_{eq}$). When $\nu_a^{\rm RS}$ becomes $\nu_a^{\rm FS}$, the lightcurve index changes from a positive value (rising lightcurve) to a negative value (falling lightcurve), resulting in a peak. 

We see that the sequence of spectral breaks in both RS and FS remain the same as the previous case. But the observed frequency this time is much lower and is still optically thick ($\nu_{\rm obs} < \nu_a^{\rm RS} = \nu_a^{\rm FS}$) unlike the previous case. Since $\numRS < \nuaRS$, the condition $1 < \left(\taumRS\right)^{1/\mu}$  holds in this case too. 

Since $\nuaFS < \numFS$, Eq-\ref{eqnua} can be written for the FS as $\nuaFS = \numFS \left(\taumFS\right)^{3/5}$. Equating this to $\nuaRS = \numRS \left(\taumRS\right)^{1/\mu}$, the spectral conditions finally lead to,
\be
1 < \left(\taumRS\right)^{1/\mu} = \left(\taumFS\right)^{3/5} (\numFS/\numRS).
\label{eq37}
\ee
A sketch of the RS and FS spectra at $t_{eq}$ is given in Fig.9 (right panel).

In MHz frequencies, RS is not likely to be significant due to the combination of two effects. First, ${f_m}^{\rm RS}$ is a sharply decreasing function of time as opposed to ${f_m}^{\rm FS}$ (which, depending on the ambient medium, either stays constant or decays at a slower rate). As $t_{\rm eq}$ is much larger than $t_a$ (of the order of $50$ times), the RS spectrum falls well below that of FS by the time of RS peak. Second, at $t_{\rm eq}$, RS is still optically thick, which further reduces the flux.

It is possible that $t_j < t_{eq}$, but we have not considered such a case in the analytical estimates. Our numerical code, however, takes care of this effect (see Figs. 1-8).

Finally, the RS and FS fluxes can be written respectively as,
\be
F^{\rm{RS}}({\nu_{\rm obs}},t_{eq}) = {F_m}^{RS} \Lb \frac{\nuaRS}{\numRS} \Rb^{-\beta} \Lb \frac{\nu_{\rm obs}}{\nuaRS} \Rb^{5/2},
\ee
and 
\be
F^{\rm{FS}}({\nu_{\rm obs}},t_{eq}) = {F_m}^{FS} \Lb \frac{\nuaFS}{\numFS}\Rb^{1/3} \Lb\frac{\nu_{\rm obs}}{\nuaFS}\Rb^{2}.
\ee

Even in the cases where RS is not significant or the RS peak is missing from observations, Eq-\ref{eq36} and Eq-\ref{eq37} can still provide sufficient insight into the underlying physical parameters if the spectral regime can be inferred. 

In appendix-\ref{section53}, we give analytical expressions of $\chi$ in high and low radio frequencies.

\subsection{Reverse shock detectability}
We use $\chi$ to probe the parameter space to find regions where RS is prominent. For this we employ numerical calculations of $\chi$, where we follow shock dynamics continuously and hence any change in spectral regimes is accounted for automatically. 

In our analytical calculations of $\chi$ we have used approximations for the thermodynamic quantities of the RS downstream. Moreover, we have not considered jet break and spectral smoothing. This has led to some deviation from the numerical values, however, the overall trend (variation with respect to physical parameters) remain the same. 

Figures 10-13 display the variation of $\chi$ across frequencies. We can see that at lower radio frequencies, the reverse shock turns insignificant. So we focus our attention on high frequencies. 

We notice a strong negative correlation of $\chi$ and the ambient medium density. For both the ISM and the wind media, a lower ambient density results in a higher $\chi$. This is in agreement with the findings of \cite{Laskar:2013uza}, where a low $A_\star$ value of GRB130427A is needed to produce a bright reverse shock. In any case, for standard microphysics parameters, {\em unless $n$ is extremely low (e.g. $n_0 < 10^{-3}~{\rm cm^{-3}}$ for ISM or $A_* < 5\times 10^{-4}$ for wind), the reverse shock component is not supposed to outshine the forward shock emission in frequencies below $\sim 1$ GHz.} 

Apart from this density dependence, a strong positive correlation is seen between $\chi$ and $\epsilon_e$. A higher $\epsilon_e$ results in a higher $\numFS$, which for a fixed $f_m^{\rm FS}$ leads to a smaller forward shock flux in radio ($\nu_{\rm obs} < \numFS$) frequencies. Since the RS is in the spectral regime $\numRS < \nu_{\rm obs} < \nuaRS$, the RS flux is not as sensitive to $\epsilon_e$ as the FS flux, resulting in a higher $\chi$. However, $\epsilon_e$ does not seem to vary largely among bursts \citep{Santana:14,Gao:2015gqa}.
As we have already seen from the lightcurves (Section-4), $\chi$ depends very weakly on $\mathcal{R}_{\rm B}$ for both types of ambient media, and the wind model has a slightly stronger dependence on $\mathcal{R}_B$ than the ISM model. 

\section{Summary and Conclusions}
In this paper, we apply the previous results of Newtonian and relativistic reverse shock dynamics to obtain radio afterglow lightcurves for both the constant density and the wind driven ambient media. Special attention is paid on a detailed analysis of synchrotron self-absorption from both the RS and the FS regions. Major findings from our investigation can be summarized as follows:

\begin{itemize}
\item Unlike optical/IR lightcurves which typically peak at the shock crossing time, the radio RS light curves (in high frequency, e.g. GHz and above) peak at the epoch when the fireball becomes optically thin to synchrotron emission. Hence, the radio reverse shock peak is delayed with respect those in the opt/IR frequencies, and appears around $\sim 1$ day.

\item In the low radio frequency regime, the forward shocked ejecta remains optically thick even when the reverse shock ejecta becomes optically thin ($\nuaFS > \nuaRS$). This sets the effective $\nu_a$ to that of the FS and changes the lightcurve evolution from what has been described earlier in the literature. The transition from RS-dominated self-absorption to FS-dominated self-absorption results in a peak in the radio reverse shock lightcurve at low frequencies (e.g. below MHz).

\item However, due to self-absorption and due to the steep decay of $f_m^{\rm RS}$, the low frequency RS lightcurve always remain well below its FS counterpart. 

\item We estimate the ratio ($\chi$) of the forward to reverse shock flux at the peak time of the reverse shock. The reverse shock is detectable when $\chi > 1$. For typical shock microphysics parameters, the RS cannot outshine the FS below 1GHz  ($\chi<1$), unless the medium density is extremely low (e.g. $n_0 < 10^{-3}~{\rm cm^{-3}}$ for ISM and $A_* < 5\times 10^{-4}$ for wind).

\item In the optical frequencies, a higher magnetization has the power to make the optical lightcurve RS dominating. But the radio RS lightcurve does not necessarily become RS dominant with a higher ejecta magnetization. 

\item A lower ambient density or a $\epsilon_e$ favors the RS detectability.
\end{itemize}

With the improved sensitivity of the Jansky VLA, it is now possible to test the reverse shock models and infer its physical parameters with better accuracy, which in turn sheds light to the nature of the ejecta \citep{Laskar:2013uza,Urata:2014bra}. \cite{Melandri:2010rk}, \cite{Kopac:2015dia} etc. have discussed various aspects of radio reverse shock emission. \cite{Gao:2013mia} has done a complete reference of analytical reverse shock lightcurves. Future more radio data from JVLA, LOFAR, FAST, and SKA are desired to test the model predictions presented in this work.
\acknowledgments
RL acknowledges support and hospitality of ARIES, Nainital, ICTS, Bangalore and University of Nevada, Las Vegas in the course of this project. BZ acknowledges NASA NNX 15AK85G and NNX 14AF85G for support.
%
%
%
%
%
\begin{appendix}
\section{Synchrotron self-absorption frequencies}
\label{appen1}
In this appendix, we present expressions for the RS self-absorption frequencies. In cases with explicit numerical dependence on $\eta$, we have normalized $\eta$ to a typical value $300$ (i.e. $\eta_{300} = \eta/300$). Numerical factors with intricate $p$-dependences are replaced with a best fit function form. The observer time $t$ is in units of seconds. All the parameters are for the RS. For the expressions as a function of the FS microphysics parameters, one can simply replace $\epsilon_e$ and $\epsilon_{\rm B}$ by ${\cal R}_e \epsilon_{\rm e,FS}$ and ${\cal R}_{\rm B} \epsilon_{\rm B,FS}$, respectively. We note that these expressions are consistent with those derived by \cite{Gao:2013mia} up to a small discrepancy in the coefficients. 

\subsection{Thin shell ISM}

Before shock crossing, for $\nu_a < \nu_m < \nu_c$, one has
\be
\nu_a =2.4 \times 10^{14} {\rm Hz} \, t^{-33/10} \, \frac{3^{3(p+1)/10} (p-1)^{8/5}}{(p-2)} \,\frac{ E_{52}^{13/10} {\epsilon_{\rm B}}^{1/5}}{n_0^{1/2} \epsilon_e \eta_{300}^{36/5}}.
\ee

For $\nu_m < \nu_a < \nu_c$, one has
\be
\nu_a = 1.6 \times 10^{13} {\rm Hz} \, 10^{-4.7p}  t^{\frac{6p-7}{p+4}} E_{52}^{\frac{3-2p}{p+4}} n_0^{\frac{5p}{8+2p}} \eta^{\frac{6(3p-2)}{p+4}} \epsilon_e^{\frac{2(p-1)}{p+4}} {\epsilon_{\rm B}}^{\frac{p+2}{8+2p}}.
\ee

For $\nu_a < \nu_c < \nu_m$, one has
\be
\nu_a = 1.2 \times 10^{15} {\rm Hz} \, 3^{3(p+1)/10} \, t^{7/10} \, E_{52}^{3/10} n_0^{3/2} \eta_{300}^{19/5} {\epsilon_{\rm B}}^{6/5}.
\ee

After shock crossing, for $\nu_a < \nu_m < \nu_c$, one gets
\be
\nu_a =2.6 \times 10^{12} {\rm Hz} \, t^{-\frac{102}{175}} \, 3^{\frac{3(p+1)}{10}} \frac{(p-1)^{8/5}}{(p-2)} \,\frac{ E_{52}^{69/175} n_0^{71/175} \eta^{8/175} {\epsilon_{\rm B}}^{1/5}}{\epsilon_e}.
\ee

For $\nu_m < \nu_a < \nu_c$, one has
\be
\nu_a = 10^{17} {\rm Hz} (24.5 p^3 -145.6 p^2 + 305.2 p - 222.3) t^{\frac{-2(52+27p)}{35p + 140}} E_{52}^\frac{18p + 58}{35p+140} n_0^{\frac{94-p}{280+70p}} \eta^{-\frac{44+74p}{35p+140}} \epsilon_e^{\frac{2(p-1)}{p+4}} {\epsilon_{\rm B}}^{\frac{p+2}{8+2p}}.
\label{eq19}
\ee

We note that since most indices have $p$-dependences, one cannot get the correct numerical coefficient without assuming a $p$ value. Therefore, the numerical values of $\nu_a$ expressions cannot be regarded as the typical value. For example, for Eq.(\ref{eq19}), for $p \sim 2.2$, the exponents of $\eta$ and $\epsilon_{\rm B}$ are $\sim -1$ and $\sim 0.3$, respectively. Taking typical values of $\eta \sim 300$ and $\epsilon_{\rm B} \sim 10^{-3}$, the typical value goes down to $\sim 10^{14}$ Hz. The same apply for the expressions for the wind case (Sect. \ref{sec:thin-wind} and \ref{sec:thick-wind}).

\subsection{Thick shell ISM}

Before shock crossing, for $\nu_a < \nu_m < \nu_c$, one has

\be
\nu_a = 10^{17} {\rm Hz} \; 3^{3(2+p)/10} (1+z) \; \frac{(p-1)^{8/5}}{(p-2)} \; \frac{E_{52}^{3/5} n_0^{1/5} {\epsilon_{\rm B}}^{1/5} }{T_{90}^{3/5} \epsilon_e {\eta_{300}}^{8/5}} t^{-3/5}.
\ee

For $\nu_m < \nu_a < \nu_c$, one has
\be
\begin{array}{rl}
\nu_a & = 10^{14} {\rm Hz} (49.1 -26.5 p + 4p^2) \; \left(\frac{p-2}{p-1}\right)^{\frac{2 (p-1)}{p+4}} (p-1)^{\frac{2}{p+4}} \; {E_{52}}^{\frac{2}{p+4}} \; {n_0}^{\frac{p+2}{2 (p+4)}} \; \eta^{\frac{2 (p-2)}{p+4}} \\ 
& {\epsilon_e}^{\frac{2(p-1)}{p+4}} \; {\epsilon_{\rm B}}^{\frac{p+2}{2 (p+4)}} \; {T_{90}}^{-\frac{2}{p+4}} \; {(1+z)}^{\frac{4}{p+4}} \; t^{-\frac{2}{p+4}}.
\end{array}
\ee

For $\nu_a < \nu_c < \nu_m$, one has
\be
\nu_a = 6.5 \times 10^{17} {\rm Hz} \; 3^{0.3(p+1)} \frac{ {E_{52}}^{17/20} {n_0}^{19/20} (1+z)^{39/20}  {\epsilon_{\rm B}}^{6/5} }{ {T_{90}}^{17/20} \eta^{3/5} } \; t^{-0.1}. 
\ee

After shock crossing, for $\nu_a < \nu_m < \nu_c$, one has
\be
\nu_a = \frac{4.0 \times 10^{14} {\rm Hz} \; 3^{0.3(p-9)} (1+z)^{6/5} (p-1)^{8/5} \; {E_{52}}^{3/5} {n_0}^{1/5} {\epsilon_{\rm B}}^{1/5}  }{  (p-2) \, {\eta_{300}}^{8/5} \, {T_{90}}^{2/3} \, \epsilon_e}.
\ee

\subsection{Thin shell wind}\label{sec:thin-wind}

Before shock crossing, for $\nu_a < \nu_m < \nu_c$, one has

\be
\nu_a = 3.6 \times 10^{11} {\rm Hz} \,  \frac{3^{\frac{3(p+1)}{10}} (p-1)^{8/5} }{(p-2)}  \; \frac{ {(1+z)}^{13/10}  {E_{52}}^{13/10}  {\epsilon_{\rm B}}^{5/2} t^{-23/10}  }{  \sqrt{A_\star} \epsilon_e {\eta_{300}}^{26/5}}.
\ee

For $\nu_m < \nu_a < \nu_c$, one has
\be
\begin{array}{rl}
\nu_a & = 2.9 \times 10^{19} 10^{-1.4 p} \, {\rm Hz} \; \frac{(p-2)^{\frac{2 (p-1)}{p+4}} (p-1)^{\frac{12}{p+4}}   }{  (p-1)^2} \; t^{\frac{p-7}{p+4}}   {(1+z)}^{\frac{3-2 p}{p+4}} {A_\star}^{\frac{5 p}{2 (p+4)}} {E_{52}}^{\frac{3-2 p}{p+4}}  \\ 
& {\epsilon_{\rm B}}^{\frac{p+2}{2 (p+4)}} {\epsilon_e}^{\frac{2 (p-1)}{p+4}} \eta^{\frac{4 (2 p-3)}{p+4}}.
\end{array}
\ee

For $\nu_a < \nu_c < \nu_m$, one has
\be
\nu_a = 
\frac{3.3 \times 10^{22} {\rm Hz} \; {A_\star}^{3/2} {E_{52}}^{3/10} 3^{\frac{3 (p+1)}{10}} {(1+z)}^{23/10} {\epsilon_{\rm B}}^{6/5}} {t^{23/10} {\eta_{300}}^{11/5}}.
\ee

After shock crossing, for $\nu_a < \nu_m < \nu_c$, one has
\be
\nu_a =
\frac{4.4 \times 10^{12} \, {\rm Hz}  \, 3^{\frac{3 (p+1)}{10}} (p-1)^{8/5} {A_\star}^{8/7}  {\epsilon_{\rm B}}^{1/5} \eta^{48/35}   }{  (p-2) {(1+z)}^{12/35} {E_{52}}^{12/35} \epsilon_e  \; t^{23/35}}.
\ee

For $\nu_m < \nu_a < \nu_c$, one has

\be
\begin{array}{rl}
\nu_a = & 10^{19} {\rm Hz} \; \frac{(p-2)^{\frac{2 (p-1)}{p+4}}(p-1)^{\frac{12}{p+4}}} {(p-1)^2} \left( 85.4 -81 p + 19.8 p^2 \right)  {(1+z)}^{\frac{2 (3p-2)}{7 (p+4)}}  \\ 
& {A_\star}^{-\frac{5 (p-10)}{14(p+4)}} {E_{52}}^{\frac{2 (3 p-2)}{7 (p+4)}} {\epsilon_{\rm B}}^{\frac{p+2}{2 (p+4)}} {\epsilon_e}^{\frac{2 (p-1)}{p+4}} \eta^{-\frac{8 (3 p-2)}{7 (p+4)}} \; t^{\frac{-13 p-24}{7 (p+4)}}.
\end{array}
\ee


\subsection{Thick shell wind}\label{sec:thick-wind}

Before shock crossing, for $\nu_a < \nu_m < \nu_c$, one has

\be
\nu_a = 3.4 \times 10^{14} {\rm Hz} \, 3^{\frac{3(p+1)}{10}} \frac{(p-1)^{8/5}}{(p-2)} \, t^{-1} \, E_{52}^{2/5} A_{\star}^{2/5} {\eta_{300}}^{-8/5} \epsilon_e {\epsilon_{\rm B}}^{1/5}  T_{90}^{2/5}.
\ee

For $\nu_m < \nu_a < \nu_c$, one has
\be
\nu_a = 10^{17} {\rm Hz} \, t^{-1} E_{52}^{\frac{2-p}{2p+8}} A_{\star}^{\frac{2+p}{p+4}} \eta^{\frac{2p-4}{p+4}} \epsilon_e^{\frac{2p-2}{p+4}} {\epsilon_{\rm B}}^{\frac{2+p}{2p+8}}  T_{90}^{\frac{p-2}{2p+8}}. 
\ee

For $\nu_a < \nu_c < \nu_m$, one has
\be
\nu_a = 4.0 \times 10^{22} {\rm Hz} \, 3^{\frac{3(p+1)}{10}} t^{-2} \, E_{52}^{-1/10} A_{\star}^{19/10} {\eta_{300}}^{-3/5} {\epsilon_{\rm B}}^{6/5} T_{90}^{1/10}.
\ee


After shock crossing, for $\nu_a < \nu_m < \nu_c$, one has
\be
\nu_a = 4.0 \times 10^{14} {\rm Hz} \,  3^{\frac{3(p+1)}{10}} \frac{(p-1)^{8/5}}{(p-2)}  t^{-3/5} \, E_{52}^{2/5} A_{\star}^{2/5} {\eta_{300}}^{-8/5} \epsilon_e  {\epsilon_{\rm B}}^{1/5}  T_{90}^{-4/5}. 
\ee

For $\nu_m < \nu_a < \nu_c$, one has
\be
\nu_a = 10^{17} {\rm Hz} \, t^{-\frac{26p+15}{8p+32}} E_{52}^{\frac{2-p}{2p+8}} A_{\star}^{\frac{2+p}{p+4}} \eta^\frac{2p-4}{p+4} {\epsilon_e}^{{2p-2}{p+4}} {\epsilon_{\rm B}}^\frac{2+p}{2p+8}  {T_{90}}^\frac{11p-14}{8p+32}.
\ee

\section{Analytical expressions of $\chi$}
\label{section53}

In this appendix, we present $\chi$ values for a typical value $p = 2.2$ in two regimes: $\chi_h$ for the high-frequency (GHz) regime, and $\chi_l$ for the low-frequency (MHz) regime:

Thin shel ISM:
\be
\chi_h = \frac{0.17 \, \text{(1+z)}^{1.4} \text{$\mathcal{R}_{\rm B}$}^{0.01} \text{$\mathcal{R}_e$}^{0.3} \text{$\epsilon_e$}^{0.1} }{\text{$E_{52}$}^{0.25} \text{$n_0$}^{0.83} \text{$\epsilon_{\rm B}$}^{0.32} \eta^{0.63}} \left( \frac{\text{$\nu_{\rm obs}$}}{10^{9}} \right)^{1.4};
\ee

\be
\chi_l = \frac{\text{$7.2 \times 10^{-6}$} \sqrt{\text{1+z}} \text{$\mathcal{R}_{\rm B}$}^{0.01} \text{$\mathcal{R}_e$}^{0.3} \text{$\epsilon_e$}^{0.06} \sqrt{\text{$\nu_{\rm obs}$}}}{\text{$E_{52}$}^{0.07} \text{$n_0$}^{0.3} \text{$\epsilon_{\rm B}$}^{0.14} \eta^{0.63}}.
\ee

Thick shell ISM:

\be
\chi_h = \frac{10^{-3} \, \text{$T_{90}$}^{0.31}\text{(1+z)}^{1.17} \text{$\mathcal{R}_e$}^{0.26}\text{$\epsilon_e$}^{0.93} \eta^{0.04}}{\text{$E_{52}$}^{0.36}\text{$n_0$}^{0.77} \text{$\mathcal{R}_{\rm B}$}^{0.02}\text{$\epsilon_{\rm B}$}^{0.35}} \left(\frac{\text{$\nu_{\rm obs}$}}{10^9} \right)^{1.48};
\ee

\be
\chi_l =  \frac{10^{-3} \, \text{$T_{90}$}^{0.31} \text{(1+z)}^{0.19} \text{$\mathcal{R}_e$}^{0.26} \eta^{0.04}}{\text{$E_{52}$}^{0.17} \text{$n_0$}^{0.18} \text{$\mathcal{R}_{\rm B}$}^{0.02} \text{$\epsilon_{\rm B}$}^{0.16} \text{$\epsilon_e$}^{0.05}} \sqrt{\frac{\text{$\nu_{\rm obs}$}}{10^9}}.
\ee

Thin shell wind:

\be
\chi_h = 
\frac{1.22 \text{$E_{52}$}^{0.9} \text{(1+z)}^{0.9}\text{$\mathcal{R}_{\rm B}$}^{0.18} \text{$\mathcal{R}_e$}^{0.49} \text{$\epsilon_e$}^{1.16}}{\text{$A_\star$}^{1.62}\text{$\epsilon _{\rm B}$}^{0.15} \eta^{1.89}} \left(\frac{\text{$\nu_{\rm obs}$}}{10^9}\right)^{0.9};
\ee

\be
\chi_l = \frac{0.12 \sqrt{\text{$E_{52}$}} \sqrt{\text{1+z}} \, \text{$\mathcal{R}_e$}^{0.26}}{\text{$A_\star$}^{0.84} \text{$\mathcal{R}_{\rm B}$}^{0.02} \text{$\epsilon_{\rm B}$}^{0.16} \text{$\epsilon_e$}^{0.06} \eta^{1.92}} \sqrt{\frac{\text{$\nu_{\rm obs}$}}{10^9}}.
\ee

Thick shell wind:

\be
\chi_h = \frac{10^{-4} \text{$E_{52}$}^{0.4} \text{$T_{90}$}^{0.56} \text{(1+z)}^{0.4} \text{$\mathcal{R}_{\rm B}$}^{0.16} \text{$\mathcal{R}_e$}^{0.5} \text{$\epsilon_e$}^{1.13} \eta^{0.08} }{\text{Astar}^{1.18} \text{$\epsilon_{\rm B}$}^{0.17}} \left(\frac{\nu_{\rm obs}}{10^9}\right)^{0.96};
\ee

\be
\chi_l =  \frac{1.2 \times 10^{-9} \text{$T_{90}$}^{0.56} \text{$\mathcal{R}_e$}^{0.19} \eta^{0.03}}{\text{$A_\star$}^{0.26} \text{$E_{52}$}^{0.06} \text{(1+z)}^{0.06} \text{$\mathcal{R}_{\rm B}$}^{0.08} \text{$\epsilon_{\rm B}$}^{0.18} \text{$\epsilon_e$}^{0.32}} \sqrt{\frac{\nu_{\rm obs}}{10^9}}.
\ee

\end{appendix}
%
%
%
%
%
%
%

%
%
%
\begin{figure*}
\begin{center}
\includegraphics[scale=0.42]{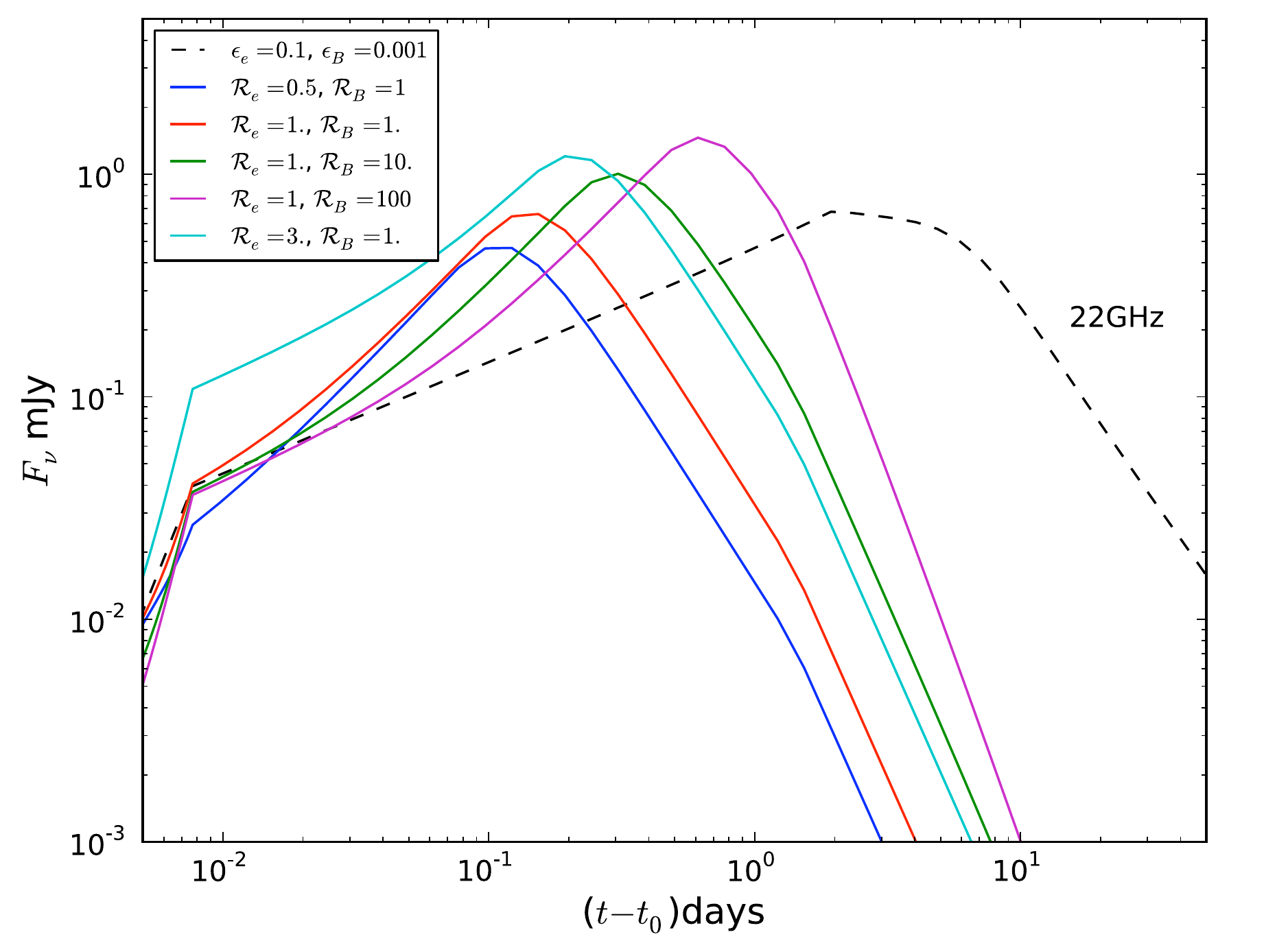}
\includegraphics[scale=0.42]{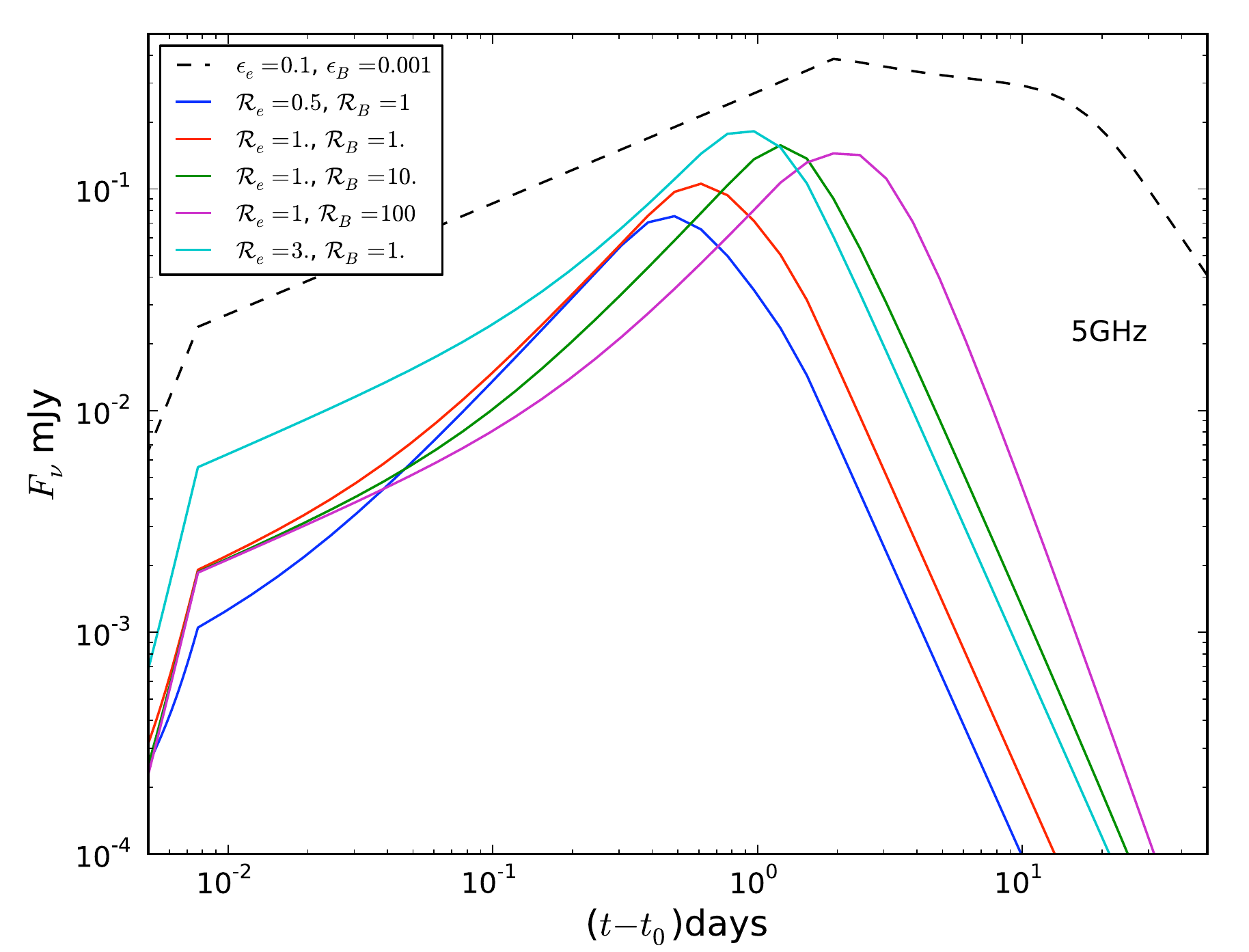}
\includegraphics[scale=0.42]{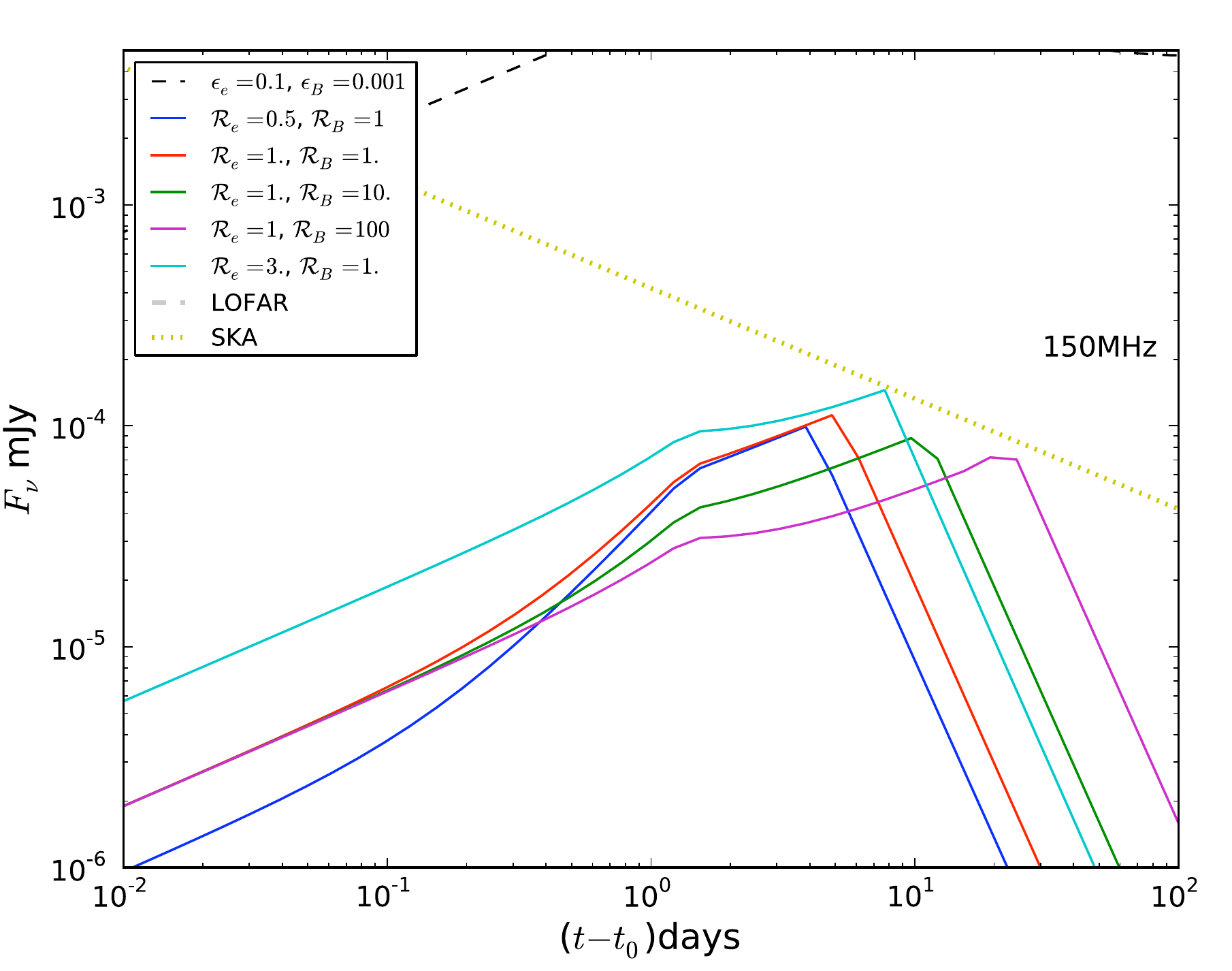}
\includegraphics[scale=0.42]{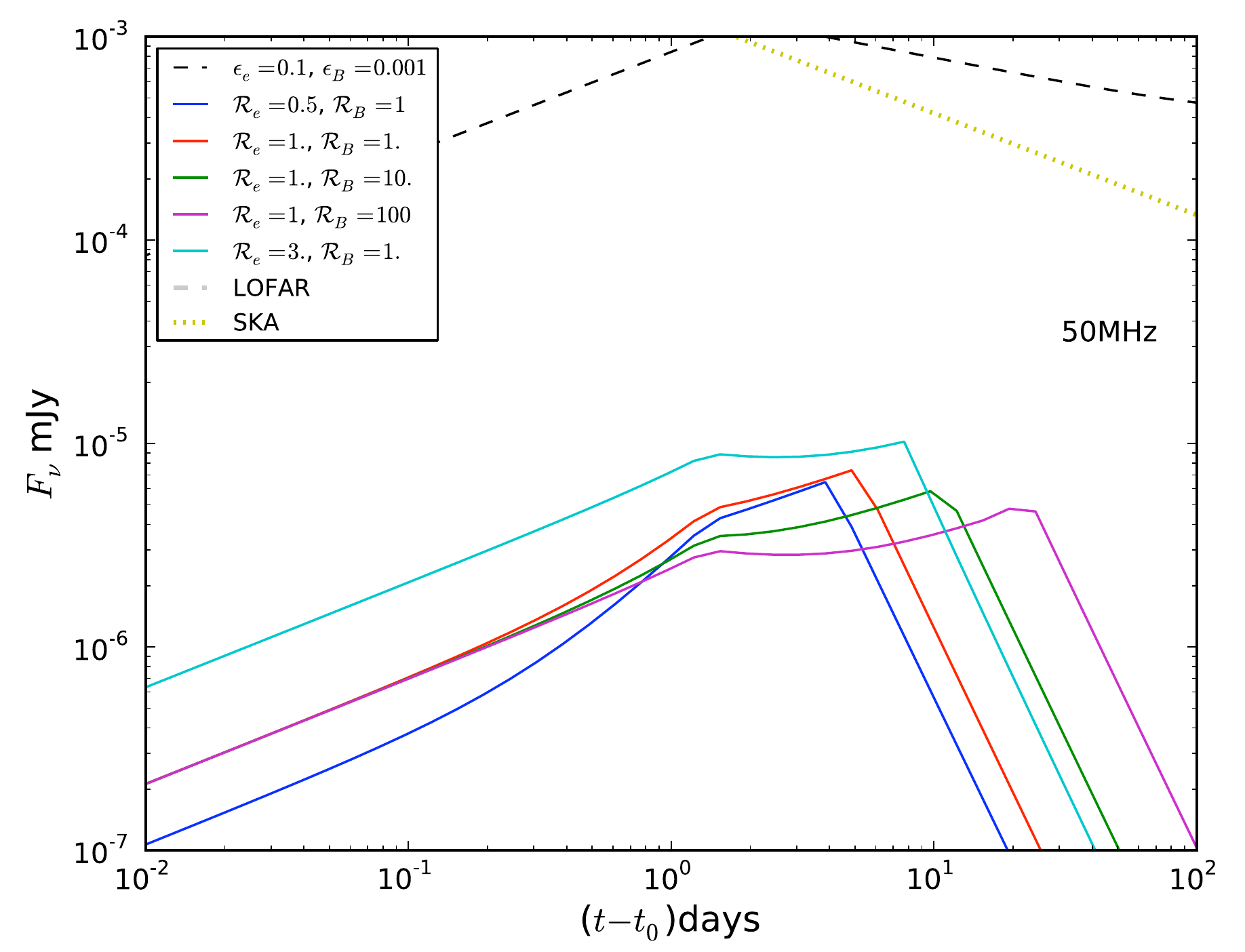}
\caption{Reverse and forward shock lightcurves for different $\mathcal{R}_e$ and $\mathcal{R}_{\rm B}$ values in the ISM model. The forward shock lightcurve is shown as dashed black curve. The physical parameters are such that the thin shell condition is satisfied : $E_{iso,52} = 5.$, $\eta = 100.$ and $n_0 = 0.1$. The redshift of the burst is assumed to be $z=1$. The RS lightcurve peak times in $22$~GHz and $5$~GHz correspond to the epochs when the fireball becomes optically thin. For $150$~MHz and $50$~MHz, the peak times are due to the change in optical thickness of the RS medium (see text). The shock-crossing time is $t_\times \sim  665$s ($0.008$~d), the transition due to which can be seen in the $22$~GHz lightcurve. The jet opening angle $\theta_j$ is $5^{\circ}$. An achromatic jet break due to the observer viewing the edge of the jet can be seen $\sim 1.5$~days in the lightcurves. For the reverse shock, the jet break suppresses the peak flux. For the forward shock, $\epsilon_e = 0.1$ and $\epsilon_{\rm B} = 0.001$ are used in all figures, and the electron distribution index $p=2.2$ is used for both forward and reverse shocks. The RS emission is calculated for a variety of ${\cal R}_e$ and ${\cal R}_{\rm B}$ values. The $5 \sigma$ sensitivity limit of the upcoming Square Kilometer Array radio telescope is shown for $150$~MHz. An integration time of $(t-t_0)/3$ and a bandwidth of $50$~MHz is used. LOFAR limits are much higher than the expected flux and the SKA limits are higher in $50$~MHz.}
\end{center}
\end{figure*}
\begin{figure*}
\begin{center}
\includegraphics[scale=0.42]{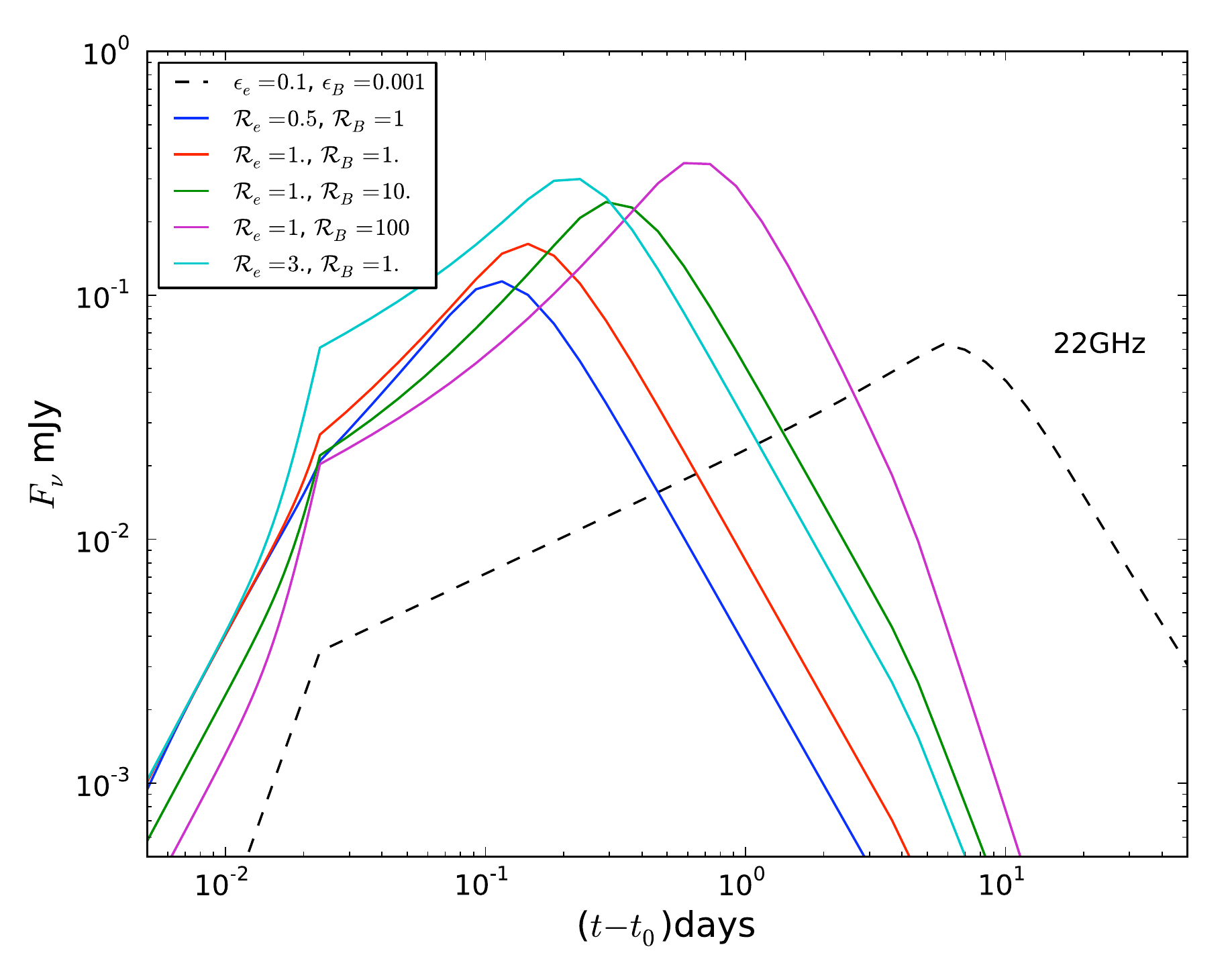}
\includegraphics[scale=0.42]{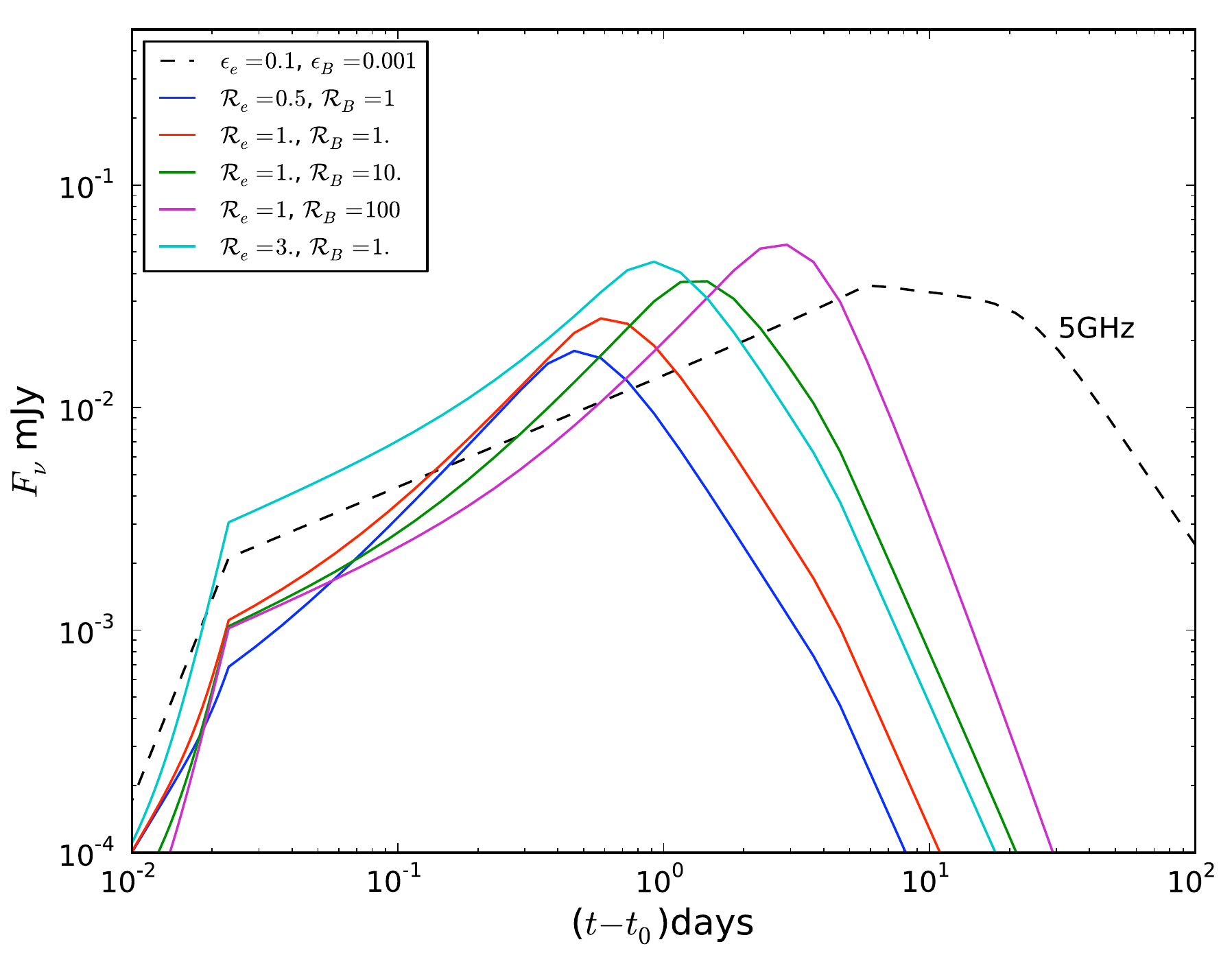}
\includegraphics[scale=0.42]{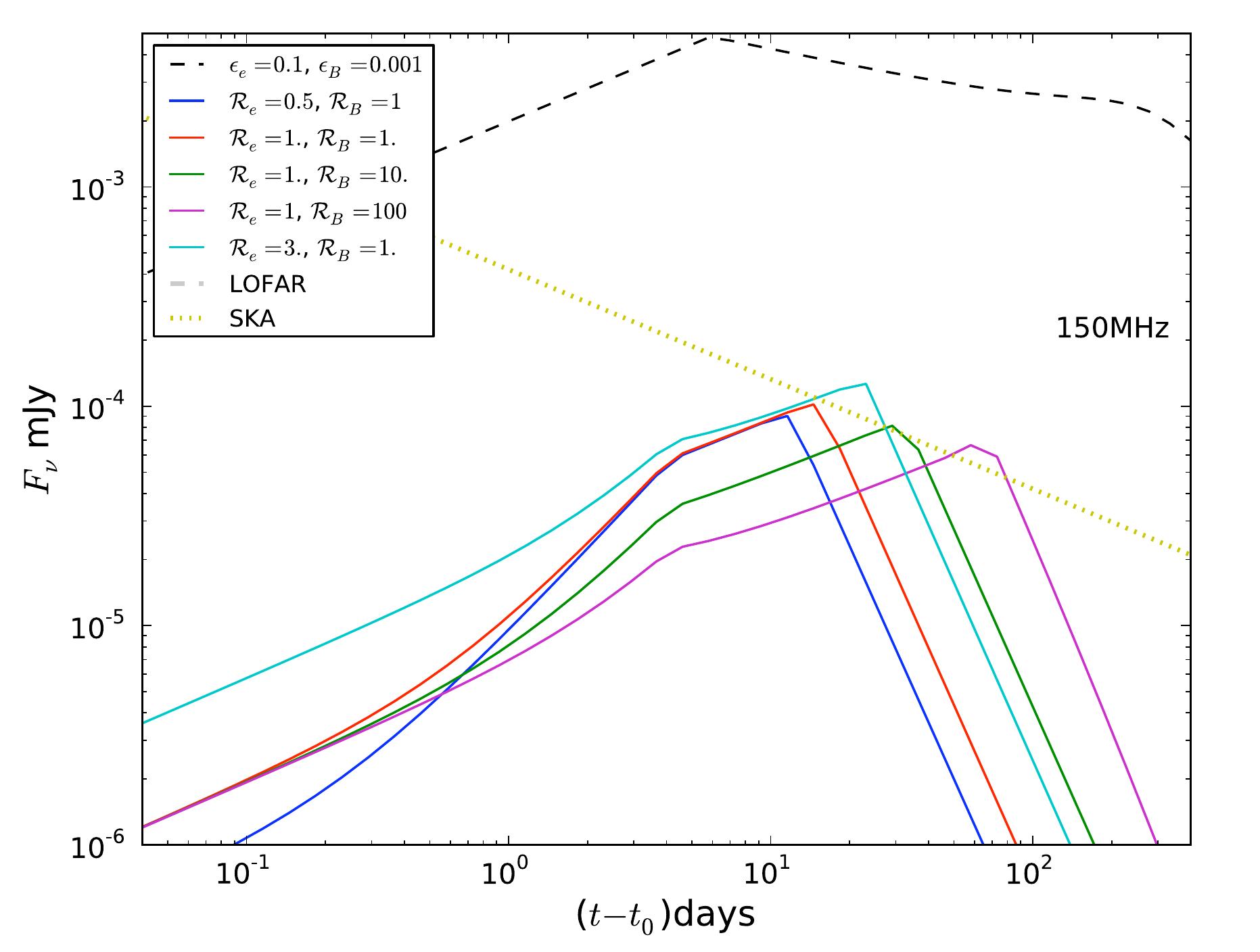}
\includegraphics[scale=0.42]{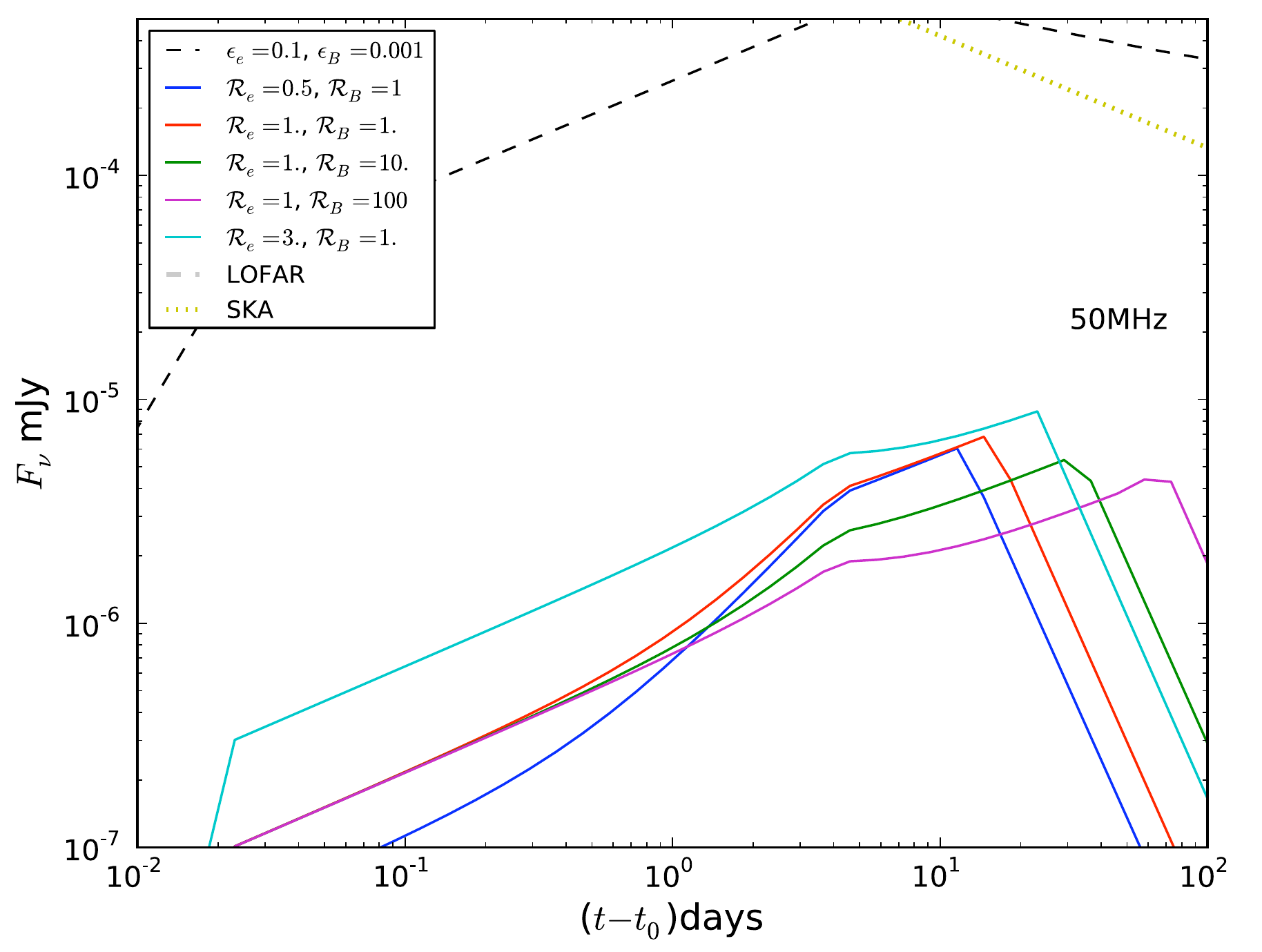}
\caption{Same as previous figure, but for $z=5$. The shock-crossing time in the observer frame increases to $t_\times \sim  2000$~s ($0.02$~d) due to time dilation.}
\end{center}
\end{figure*}
%
%
%

\begin{figure*}
\begin{center}
\includegraphics[scale=0.42]{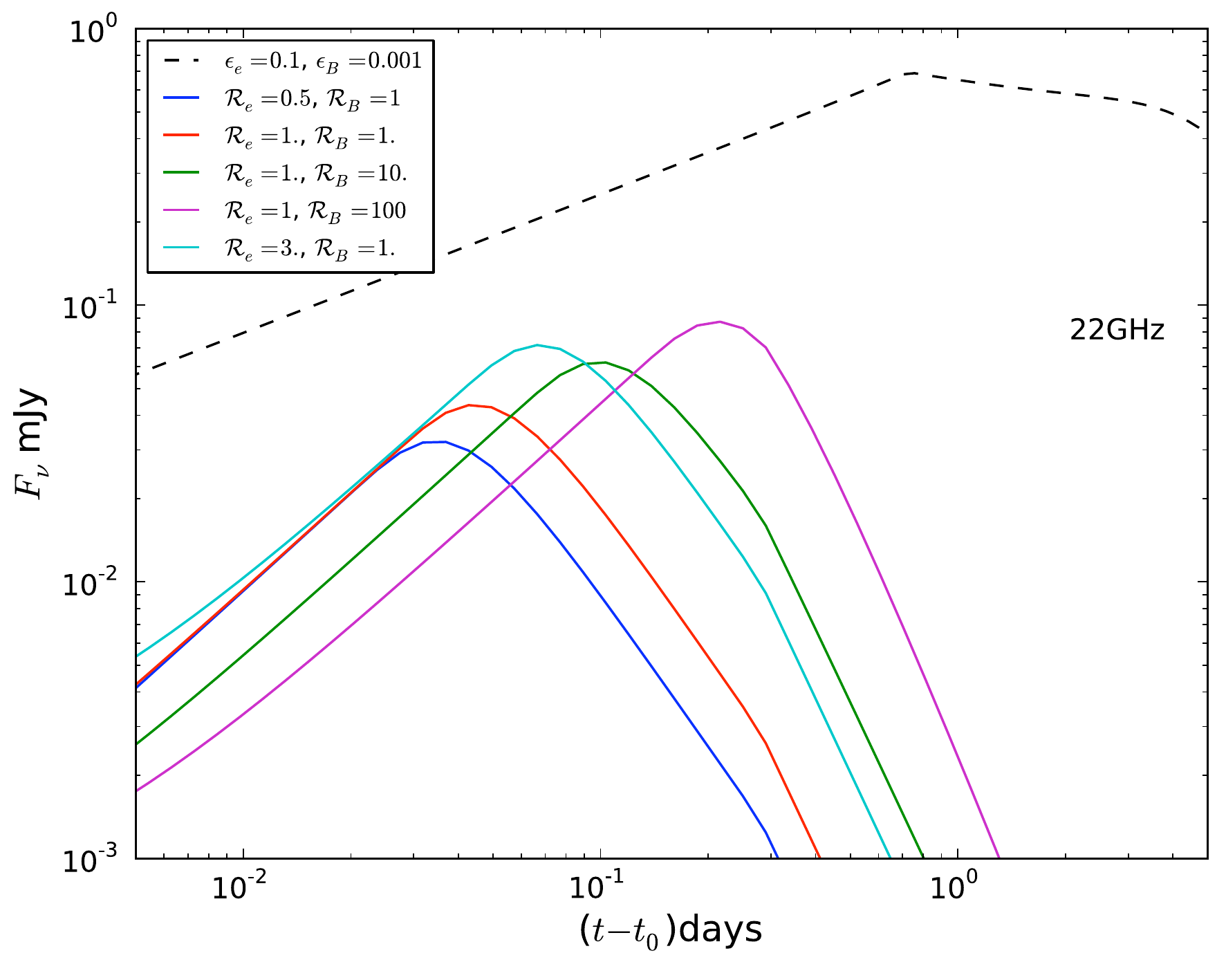}
\includegraphics[scale=0.42]{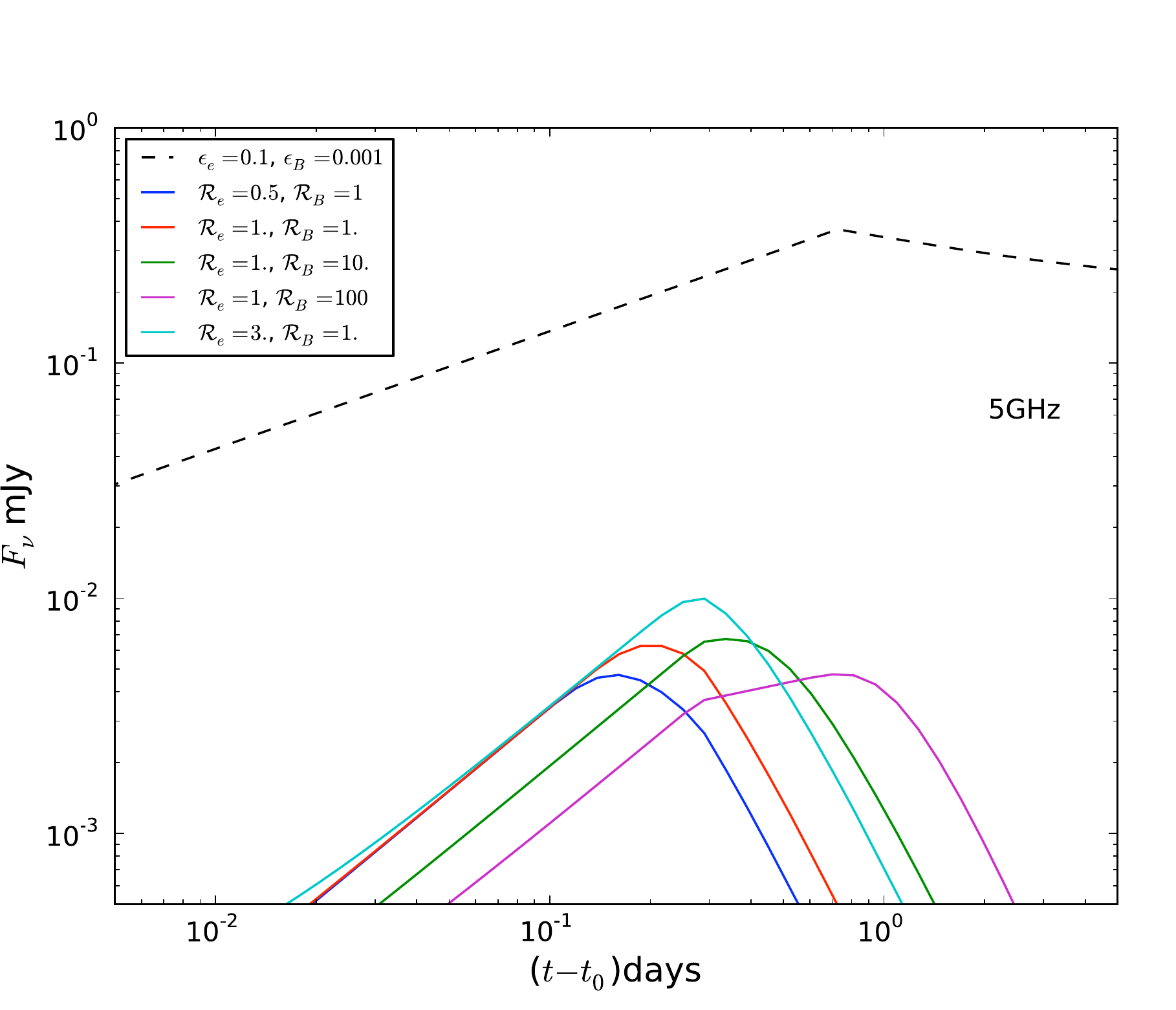}
\includegraphics[scale=0.42]{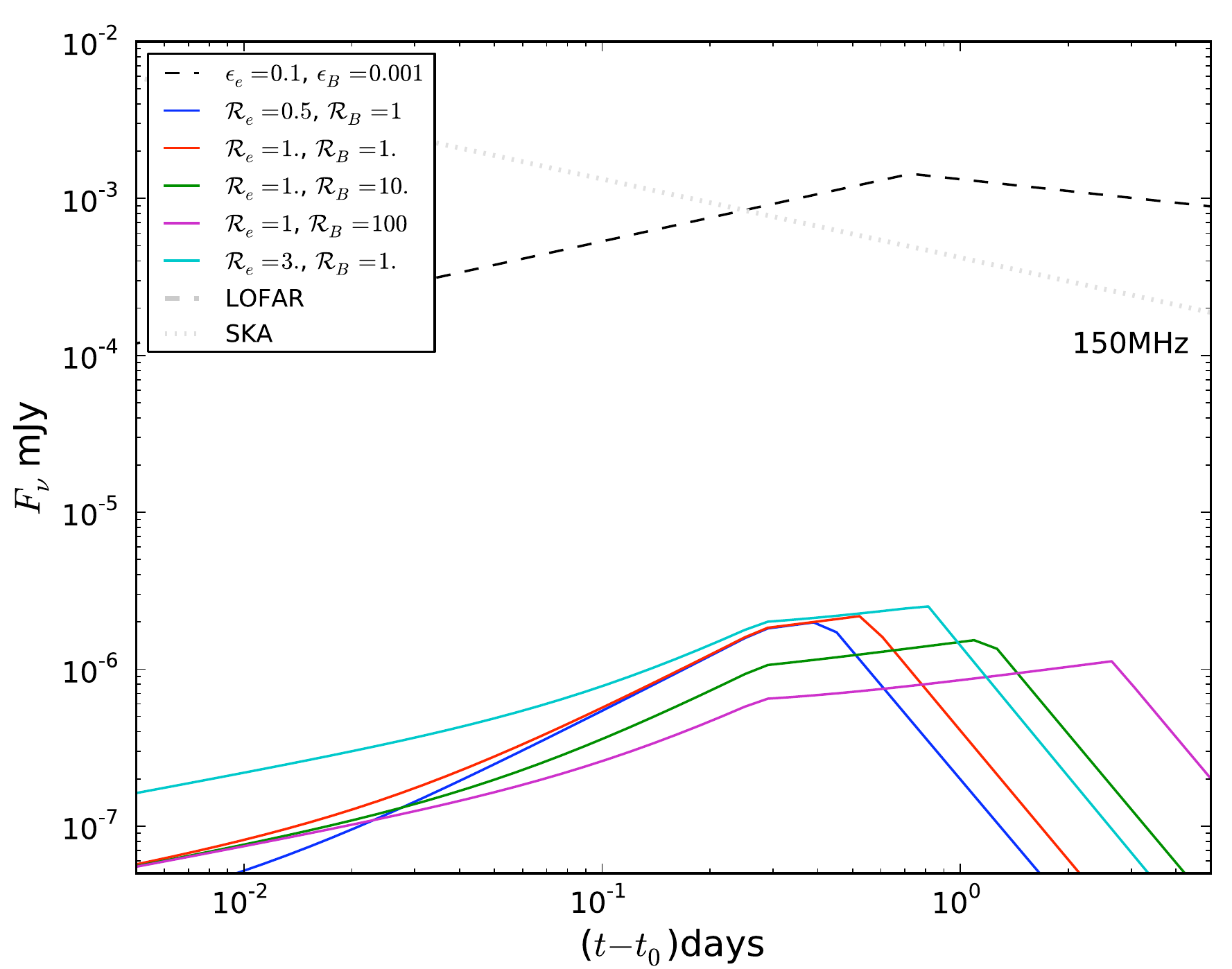}
\includegraphics[scale=0.42]{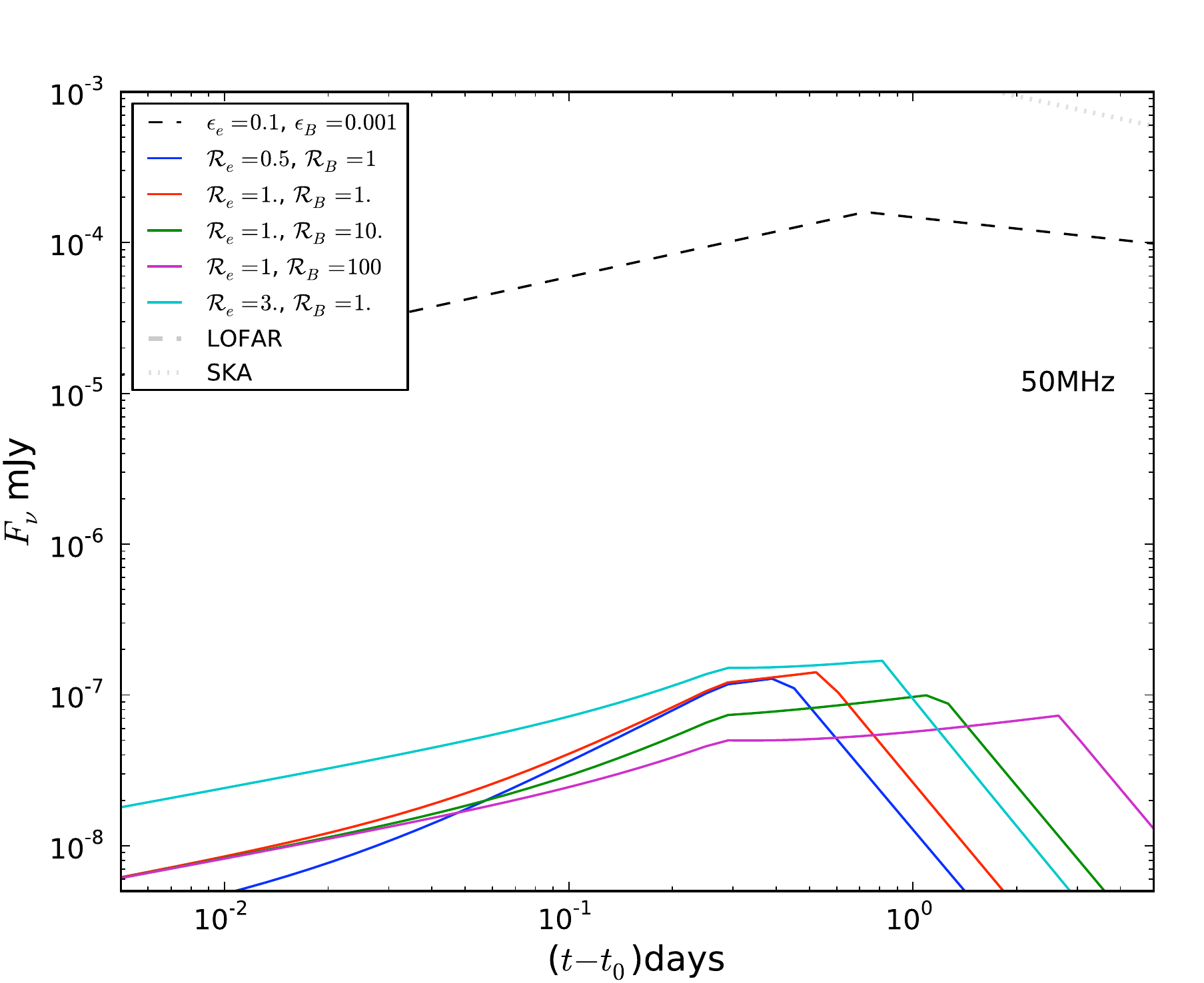}
\caption{Lightcurves for the ISM model, but the physical parameters are such that the thick shell condition is satisfied : $E_{iso,52} = 5.$, $\theta_j = 5^{\circ}$, $\eta = 400.$, $n_0 = 1.$, and $T = 80$s, $z = 1.$. The additional break before the RS peak in the high $\mathcal{R}_{\rm B}$ (magenta) case in $5$~GHz is the jet break which can be seen more prominently in the rising part of the lower frequency light curves.}
\end{center}
\end{figure*}

\begin{figure*}
\begin{center}
\includegraphics[scale=0.42]{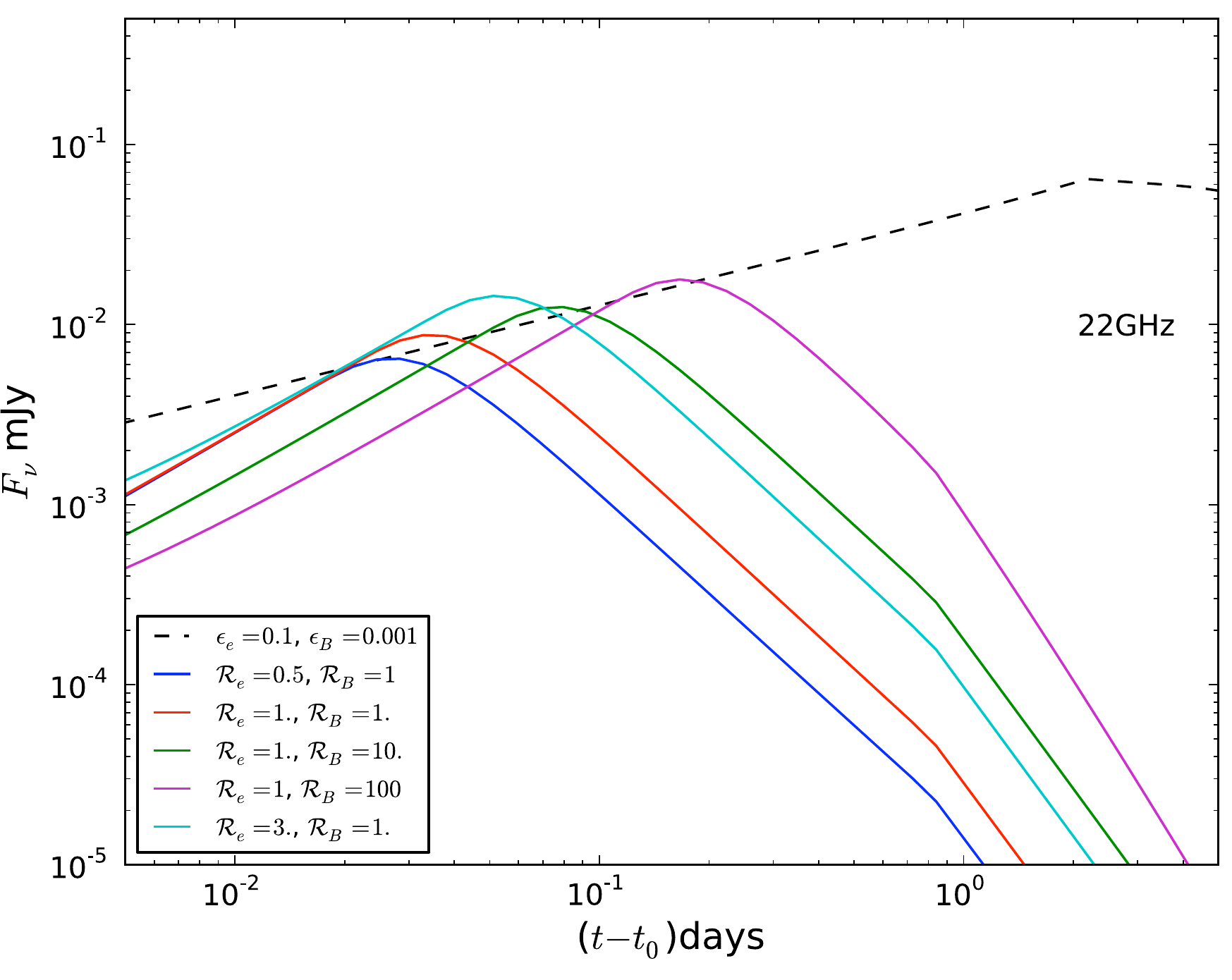}
\includegraphics[scale=0.42]{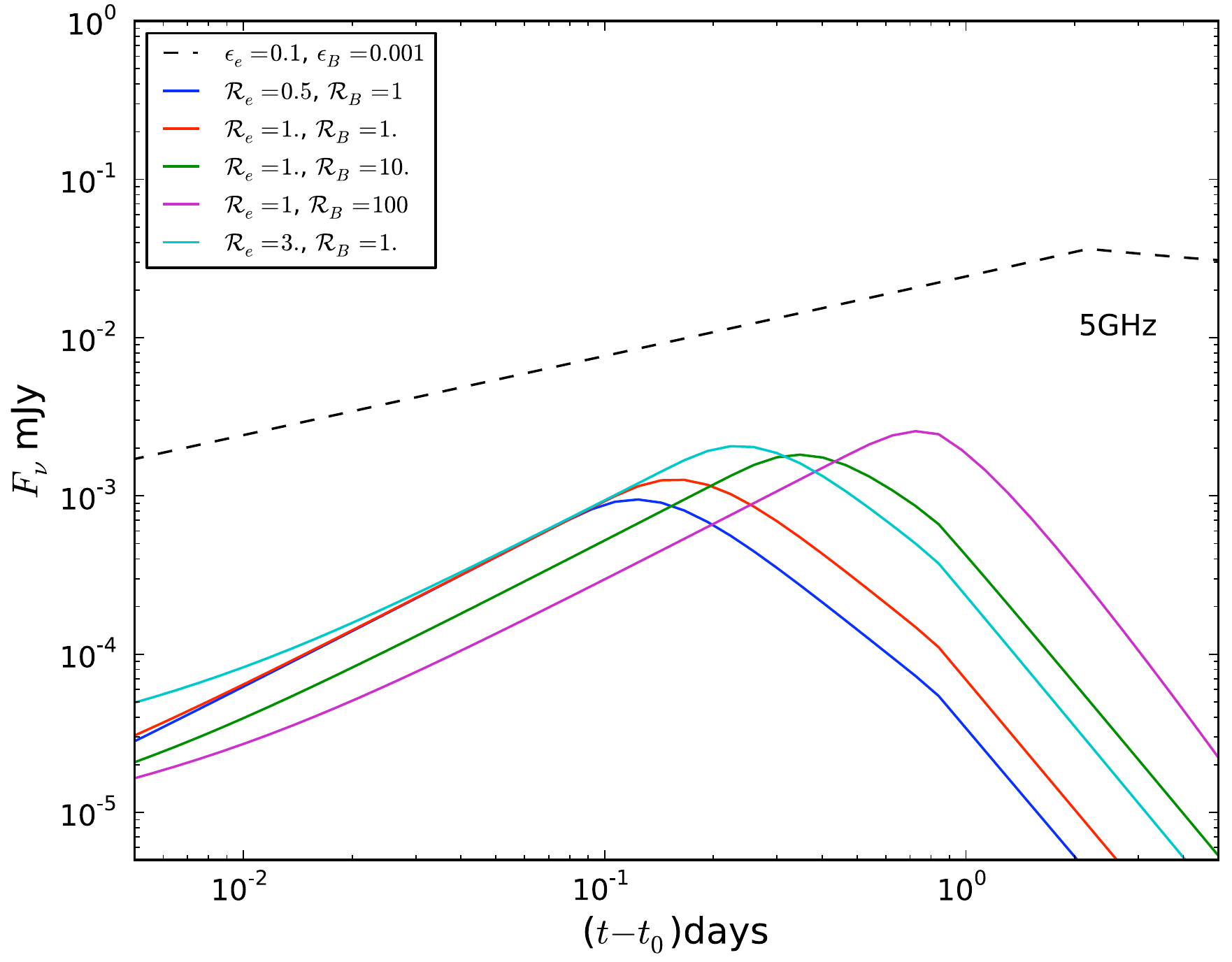}
\includegraphics[scale=0.42]{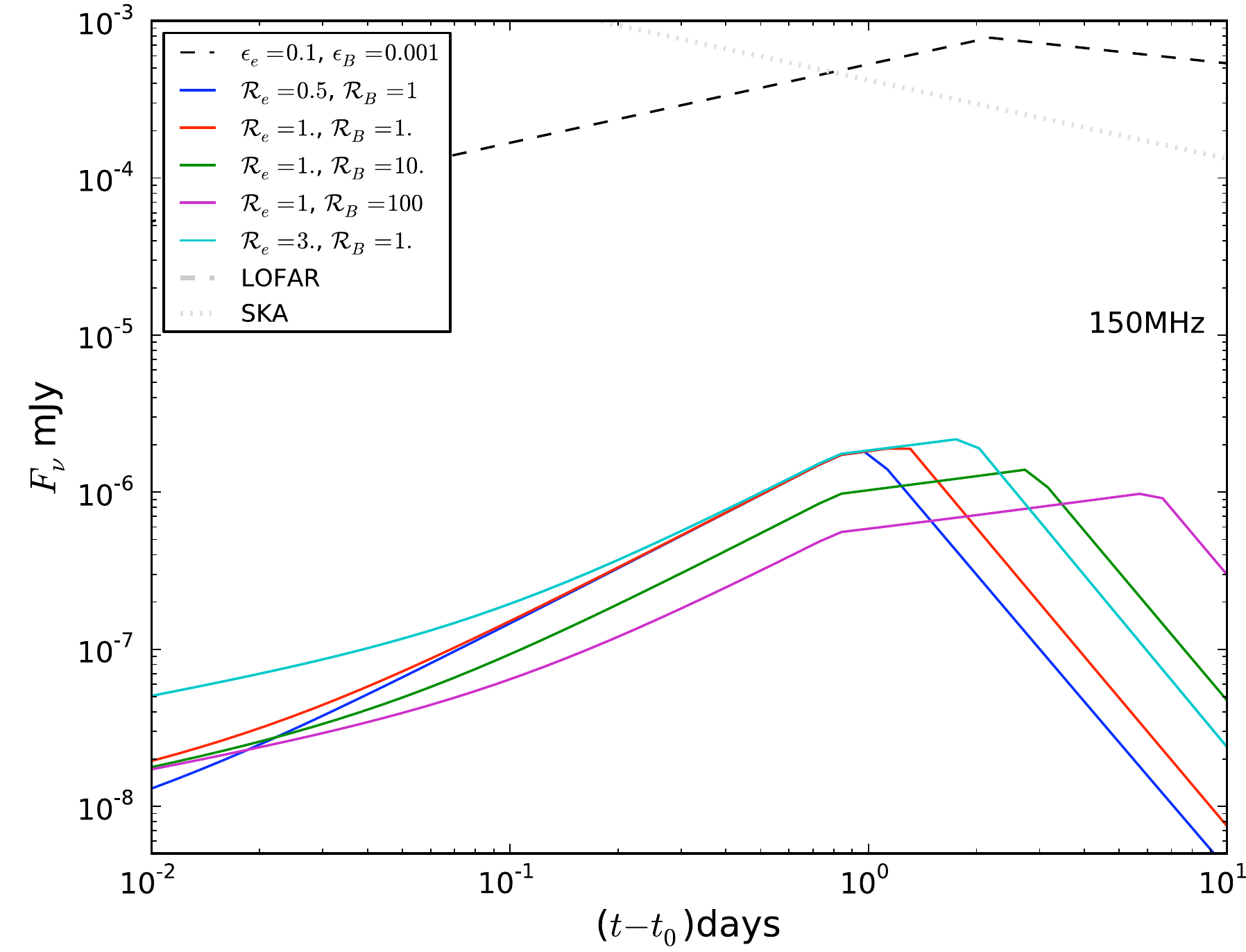}
\includegraphics[scale=0.42]{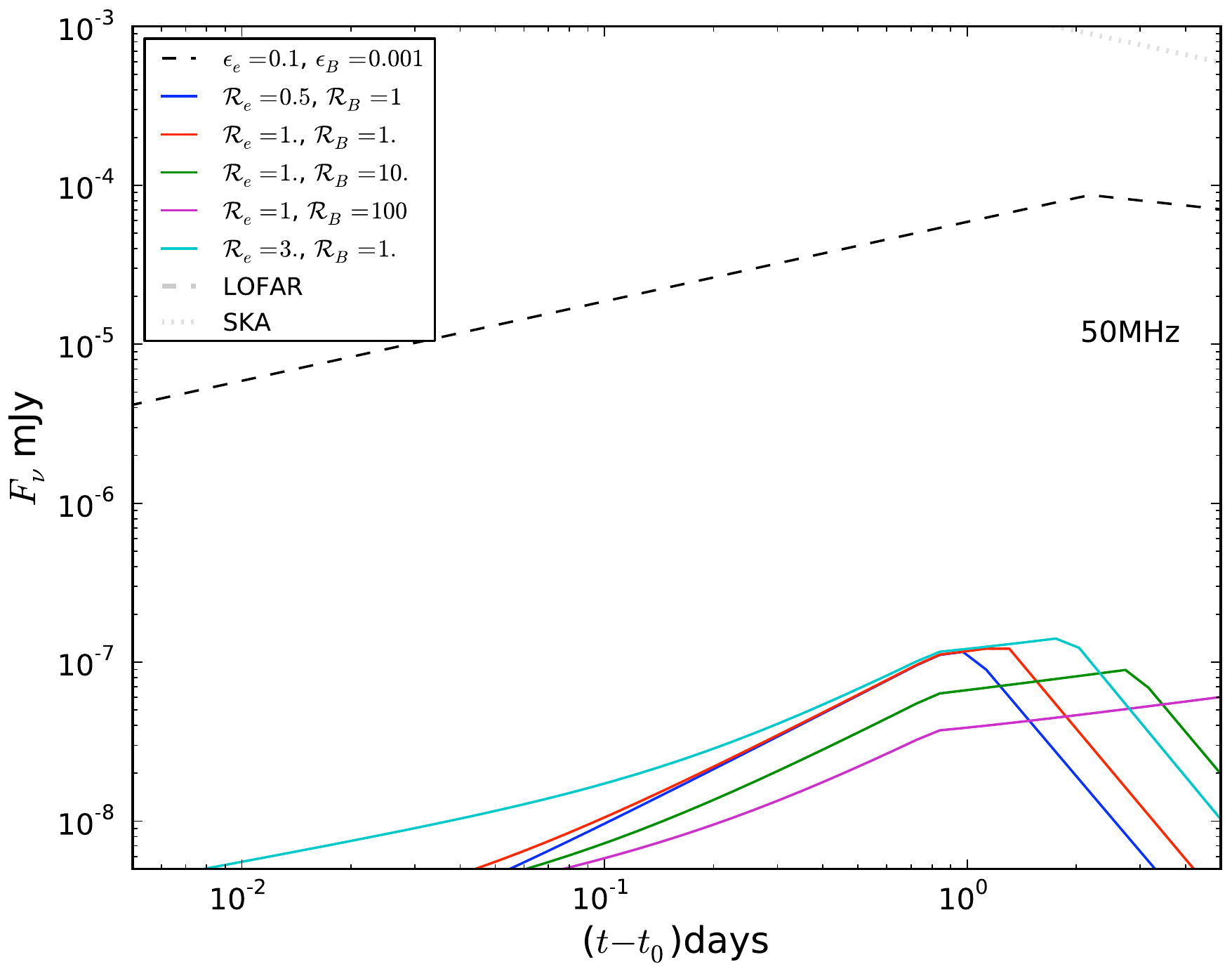}
\caption{Same as previous figure, for $z=5$.}
\end{center}
\end{figure*}

\begin{figure*}
\begin{center}
\includegraphics[scale=0.42]{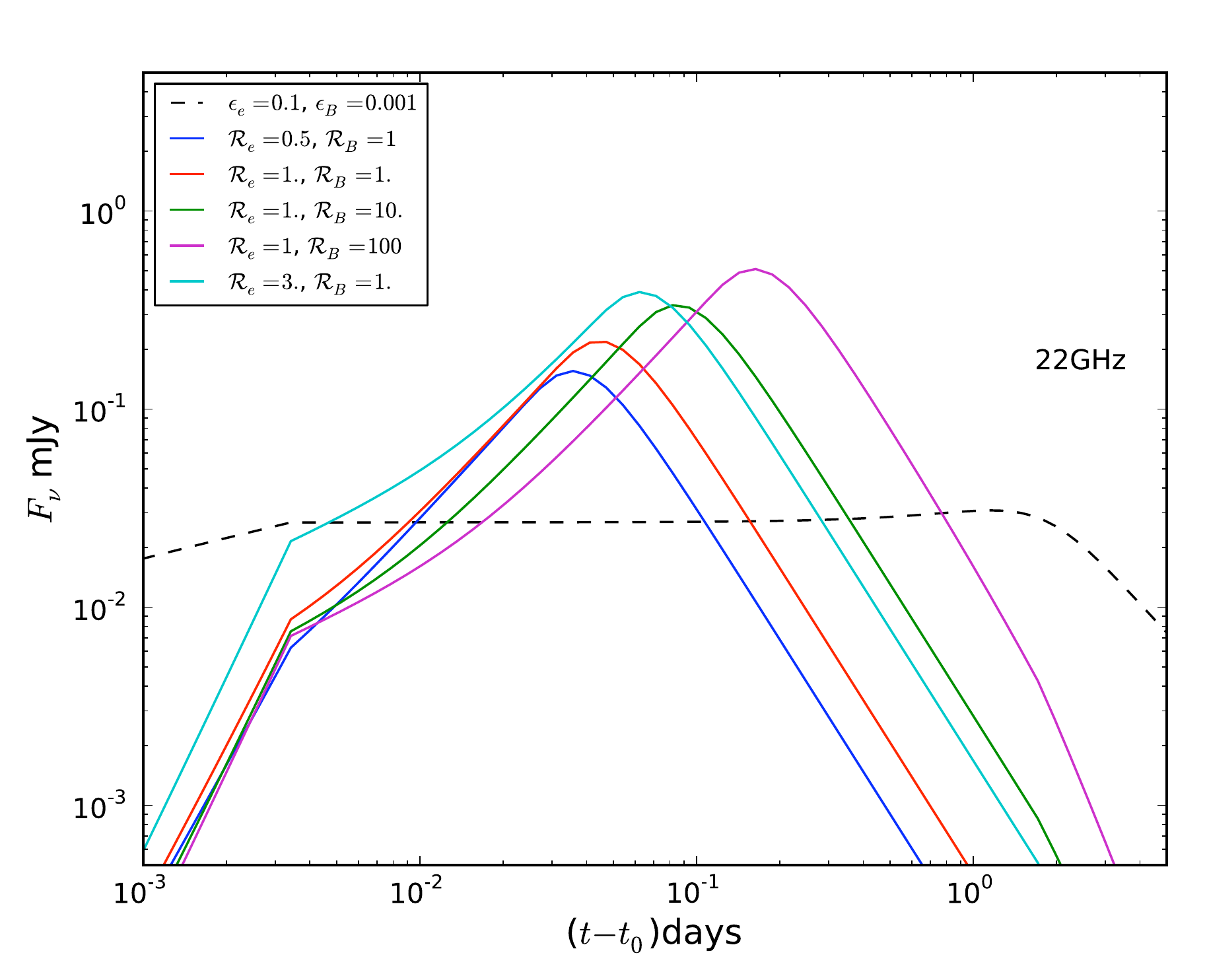}
\includegraphics[scale=0.42]{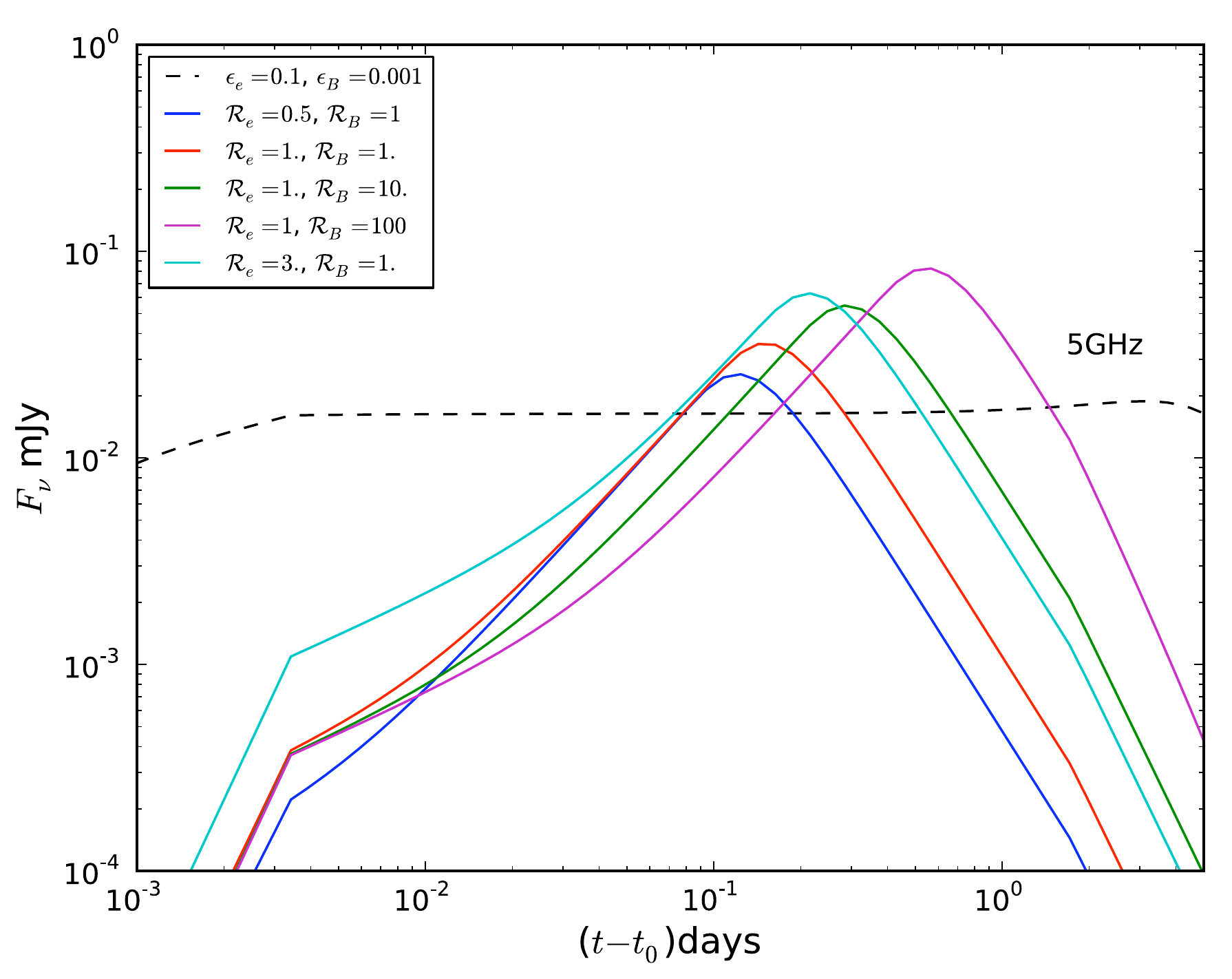}
\includegraphics[scale=0.42]{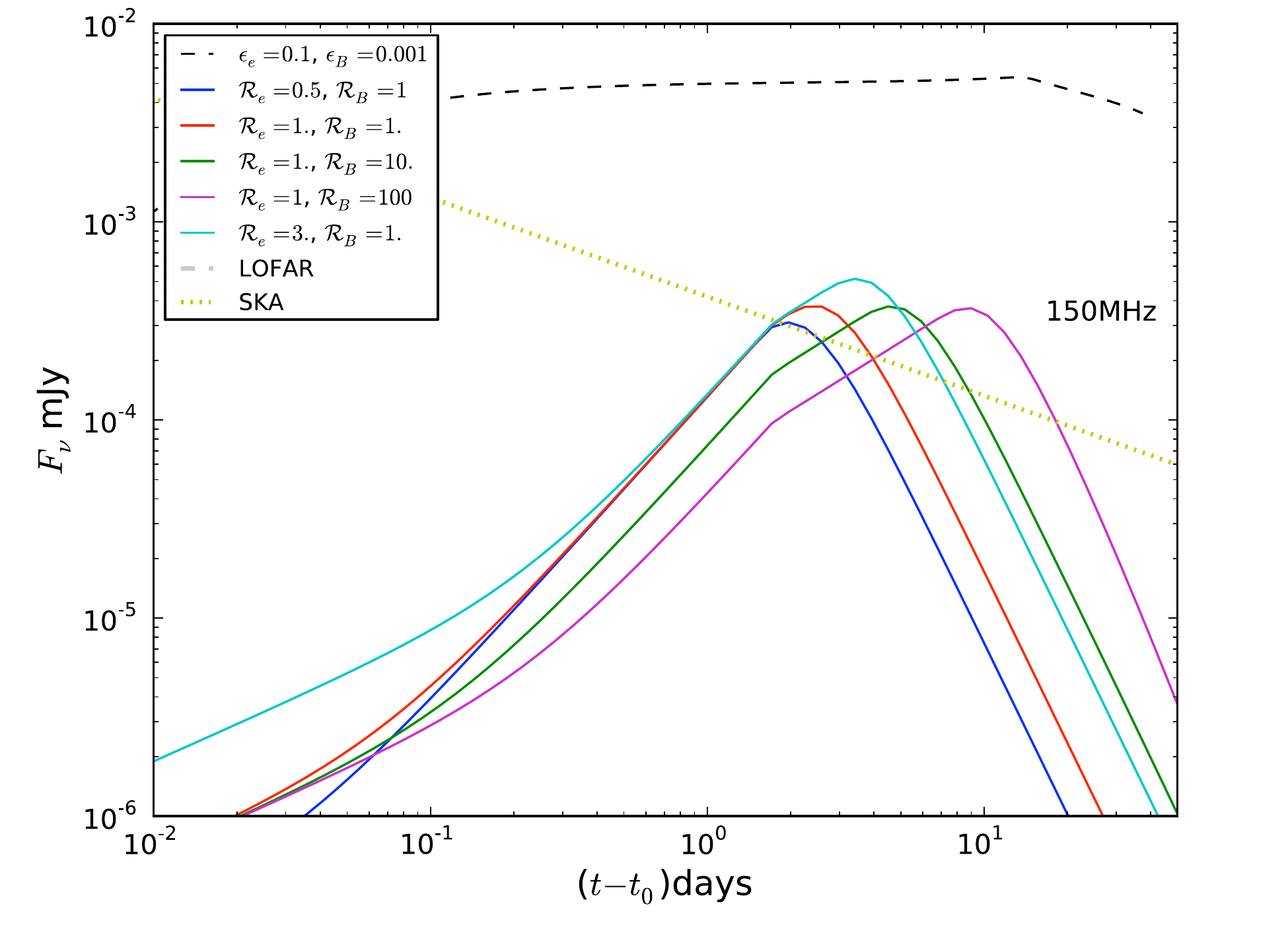}
\includegraphics[scale=0.42]{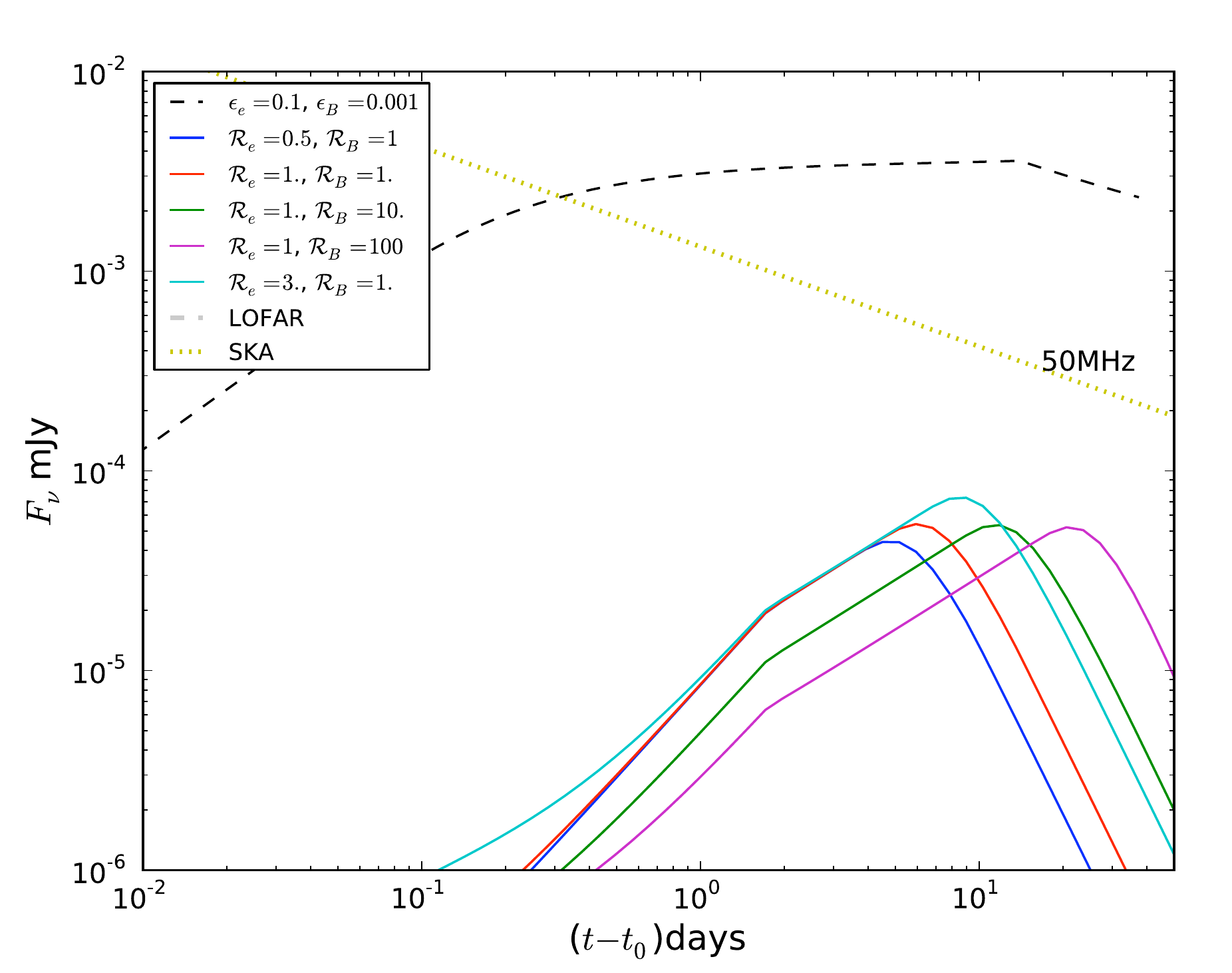}
\caption{Same as Fig.1 but for the thin shell wind model. Physical parameters are : $E_{iso,52}= 5.$, $A_{\star} = 0.01$ and $\eta = 100$, $z=1$ which leads to $t_\times =  295$s ($0.003$ day). The forward shock flux is nearly constant as both $22$~GHz and $5$~GHz are between $\nuaFS$ and $\numFS$. The physical parameters are such that unlike the previous figures, the peak time corresponds to the crossing of $\nu_a$ through the radio band including for lower frequencies.}
\end{center}
\end{figure*}
\begin{figure*}
\begin{center}
\includegraphics[scale=0.42]{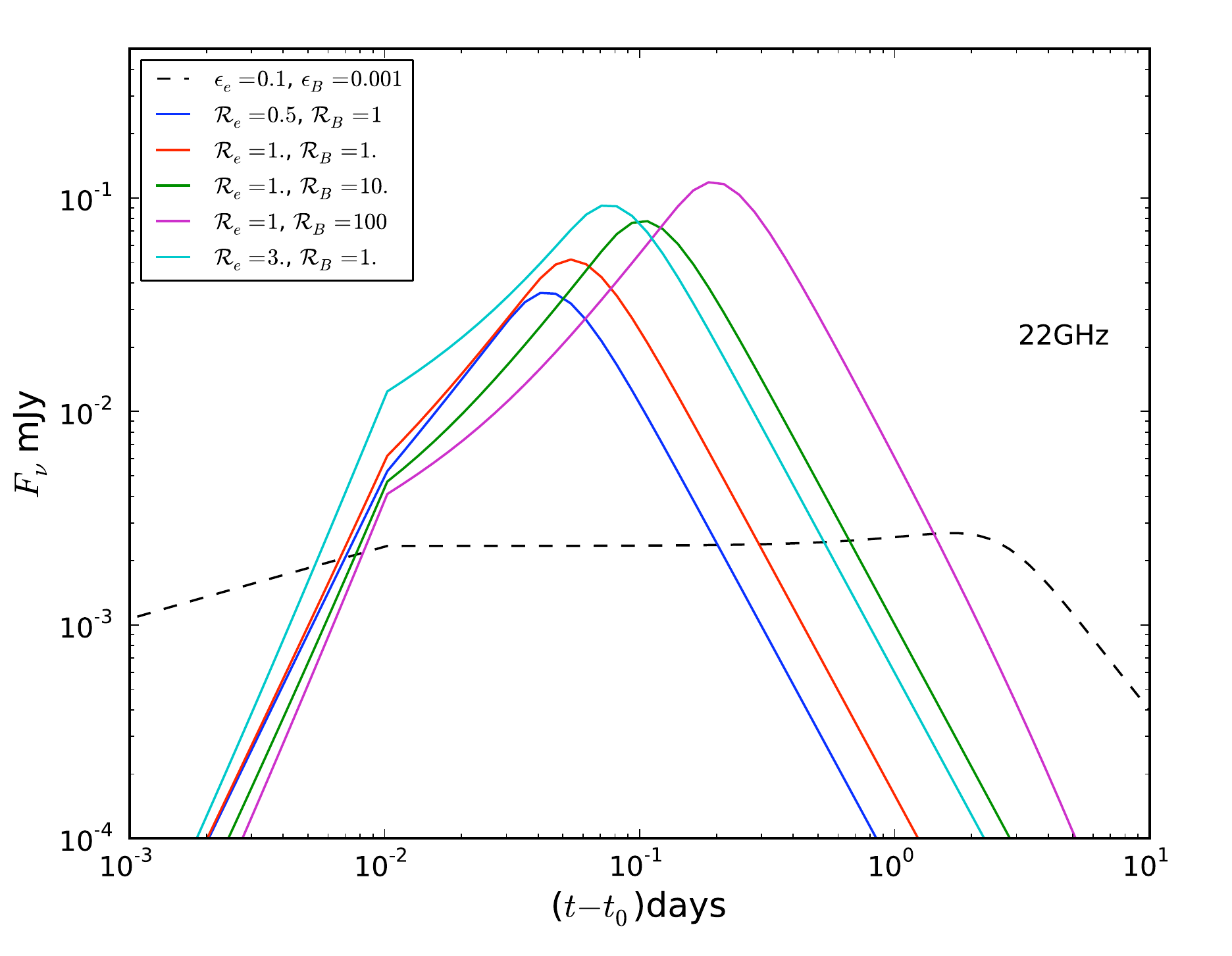}
\includegraphics[scale=0.42]{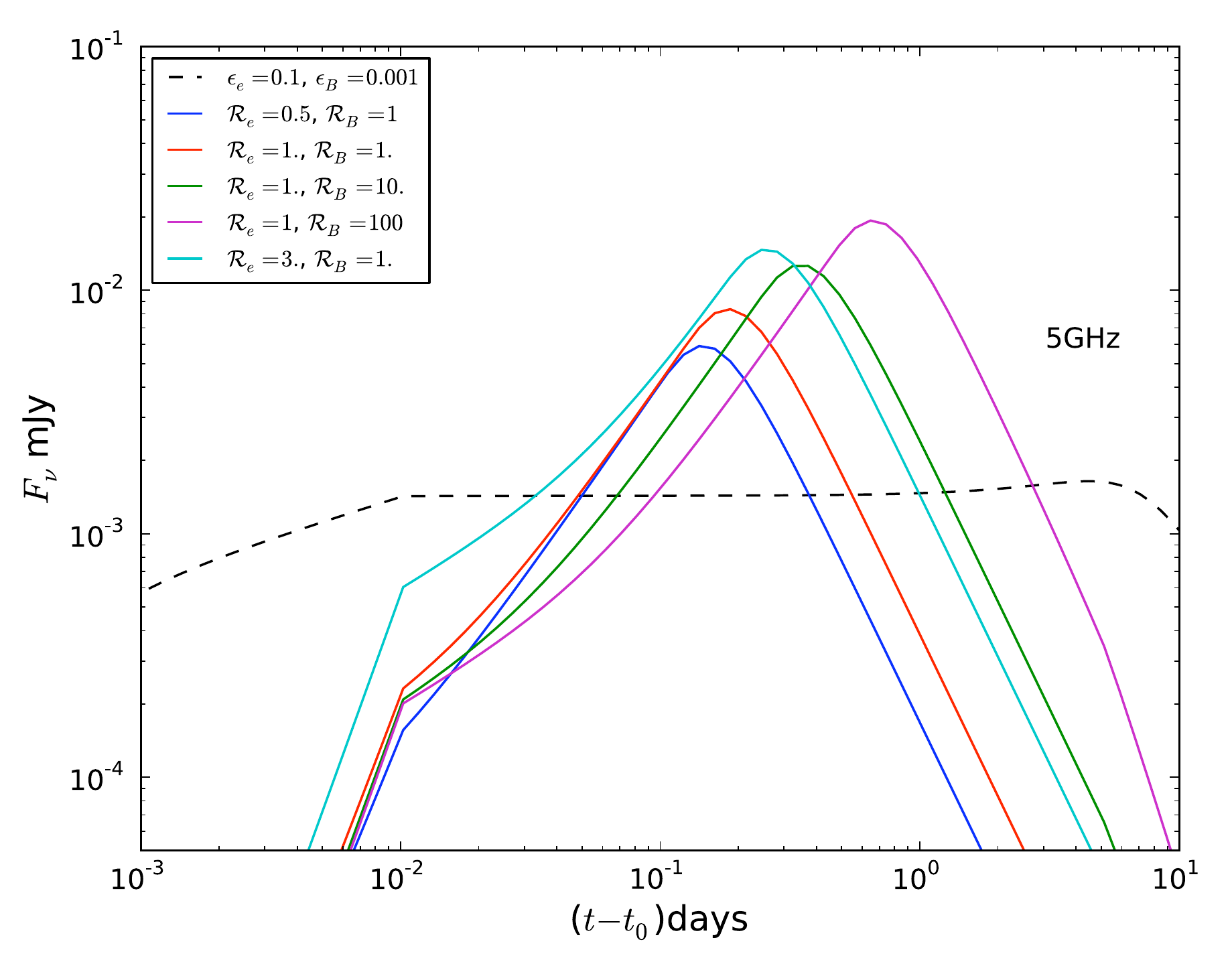}
\includegraphics[scale=0.42]{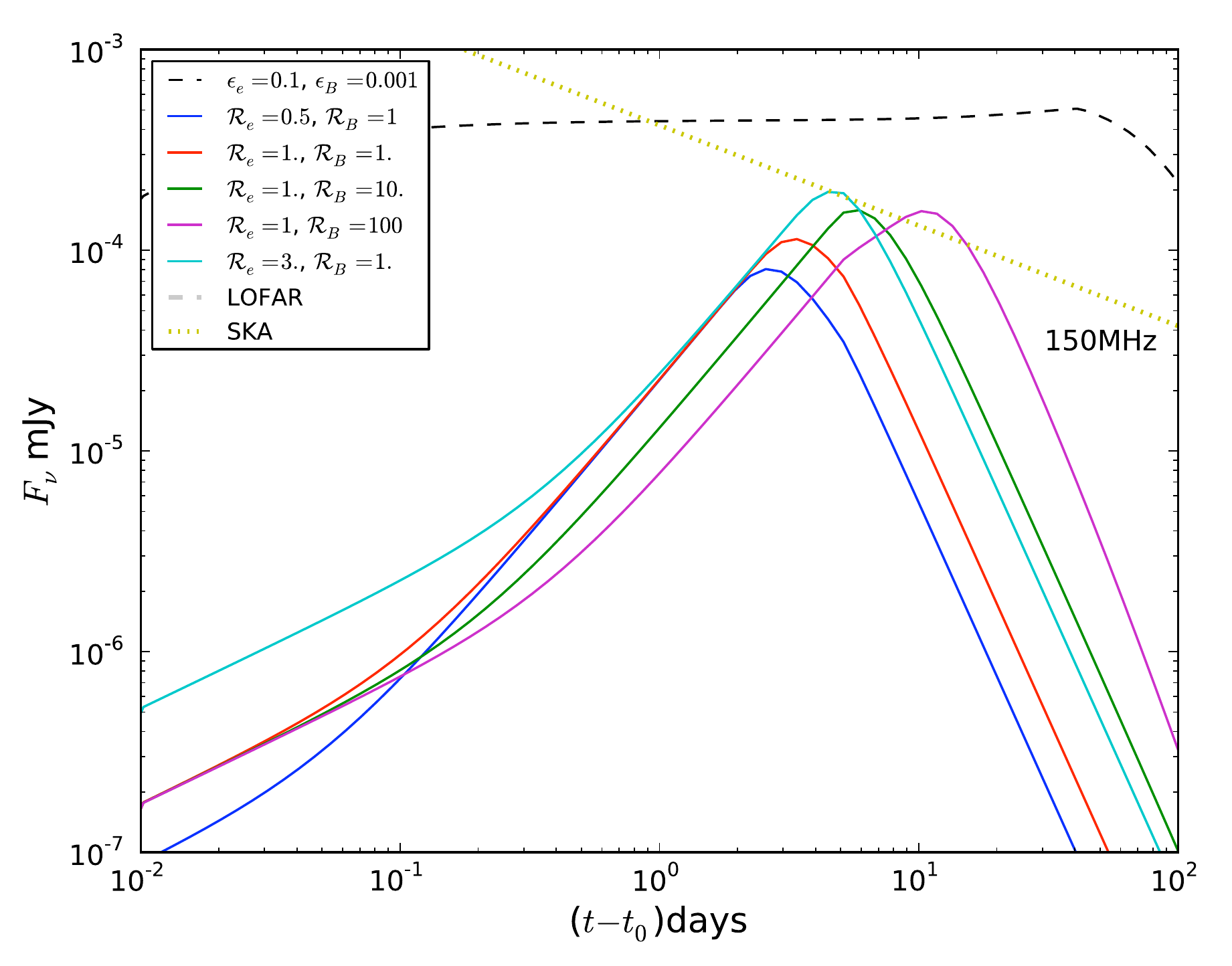}
\includegraphics[scale=0.42]{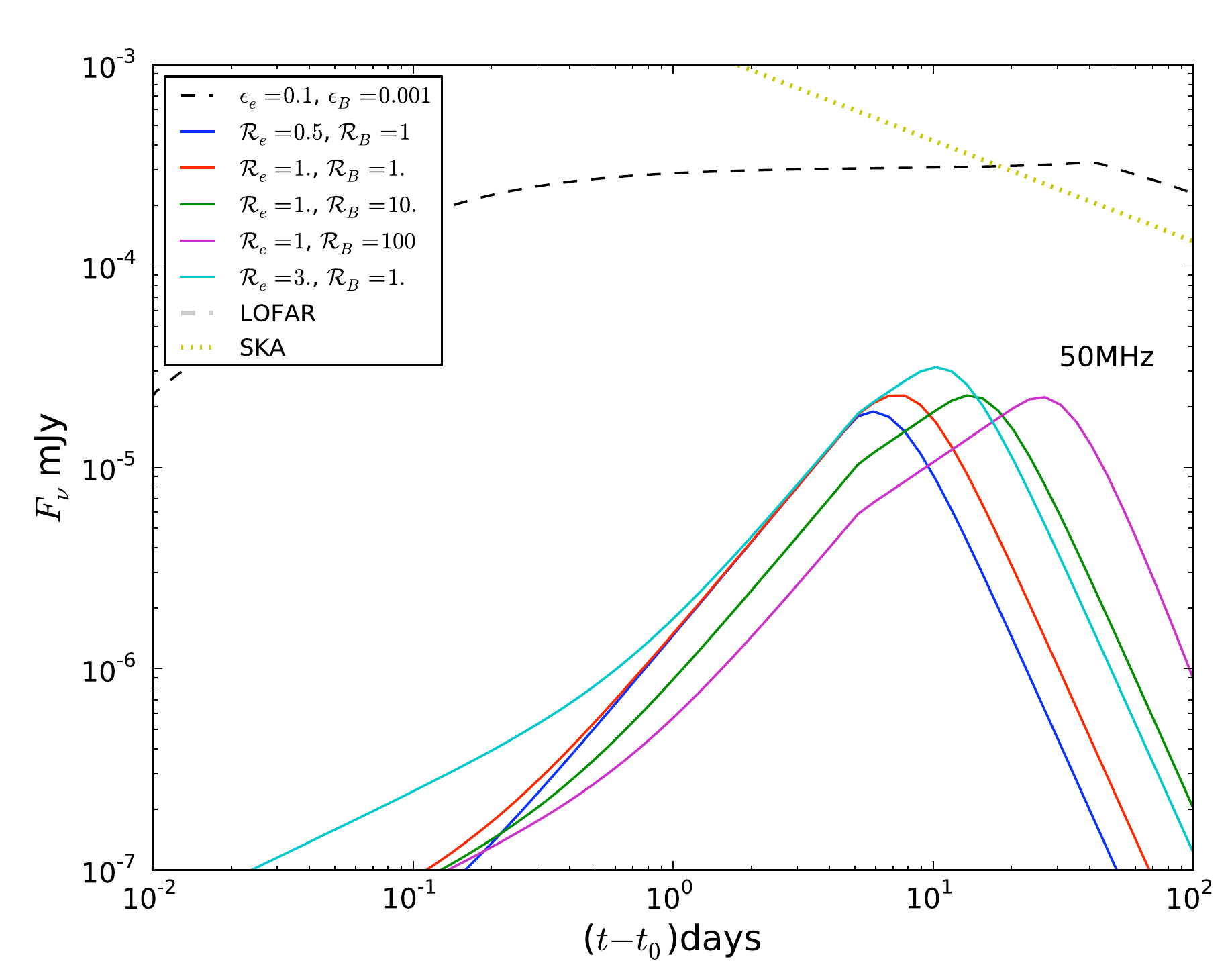}
\caption{Same as previous figure, for $z=5$.}
\end{center}
\end{figure*}
%
%

\begin{figure*}
\begin{center}
\includegraphics[scale=0.42]{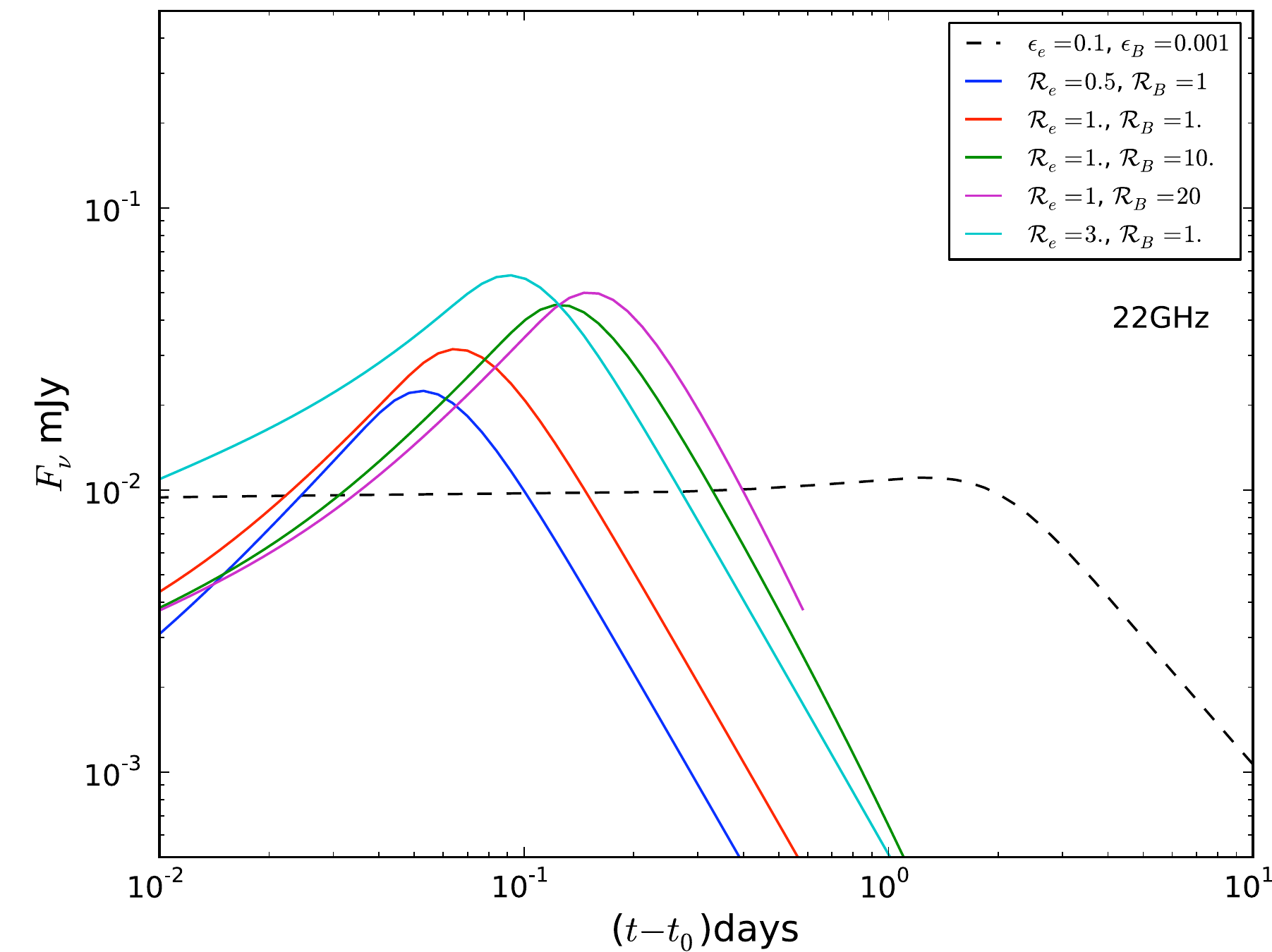}
\includegraphics[scale=0.42]{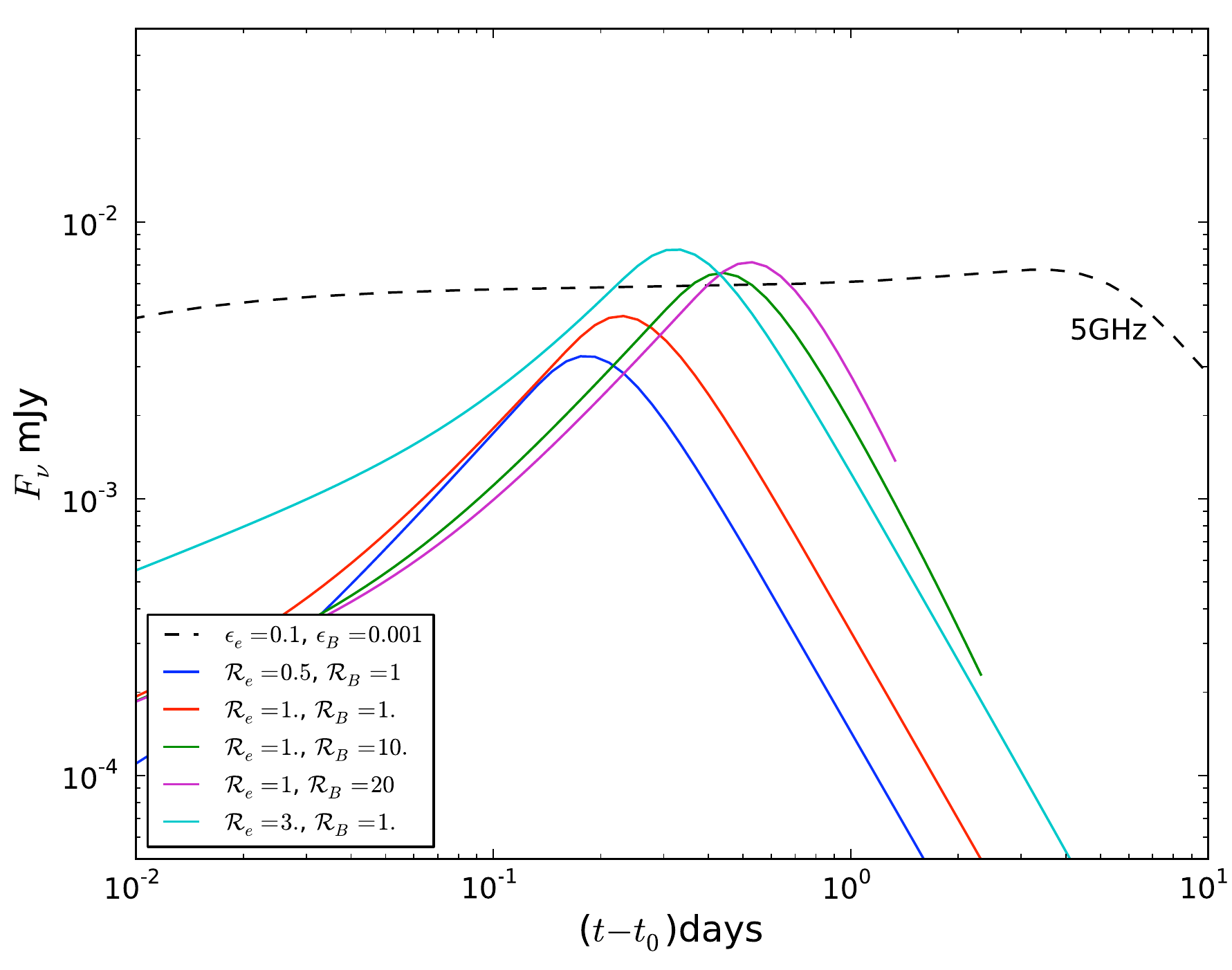}
\includegraphics[scale=0.42]{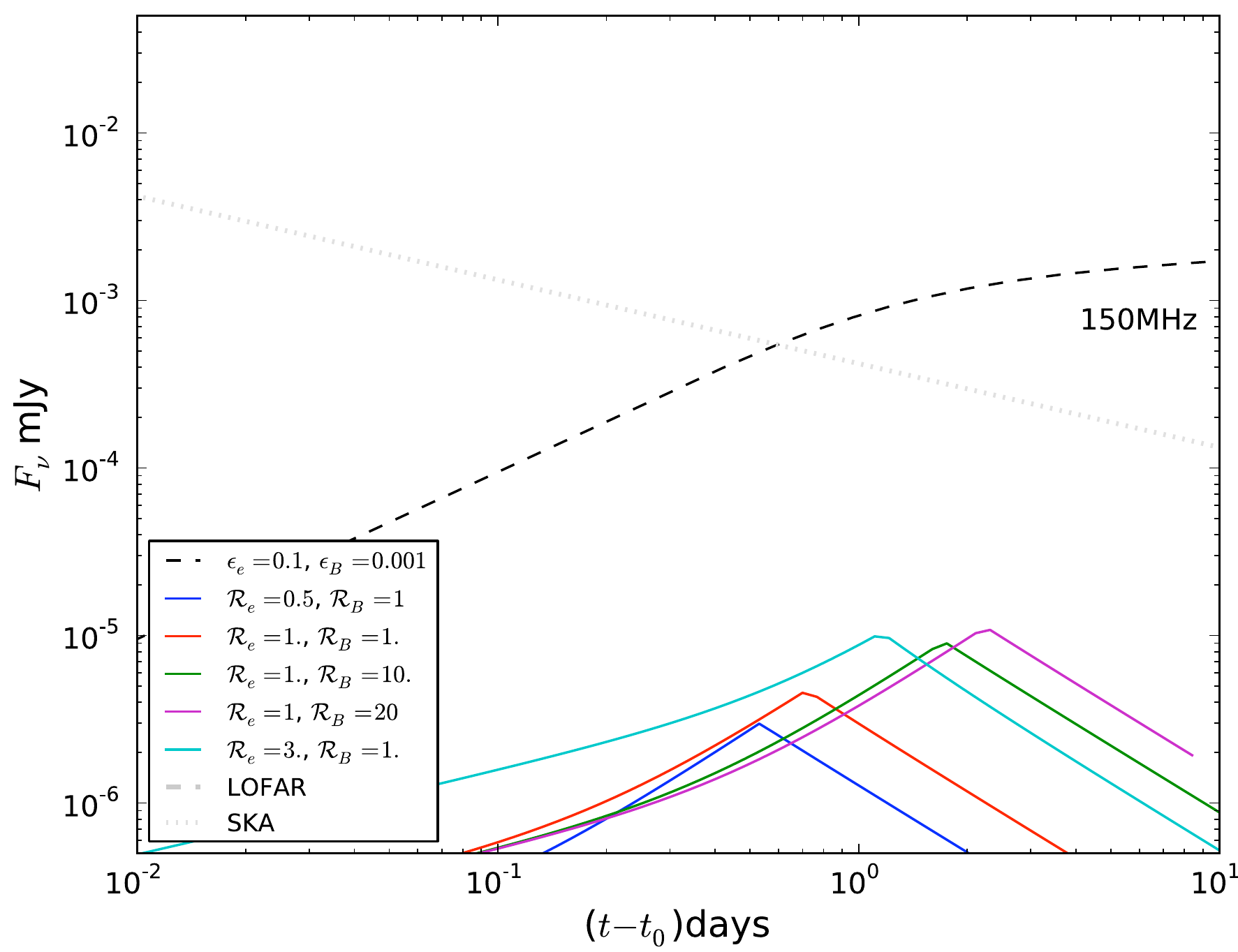}
\includegraphics[scale=0.42]{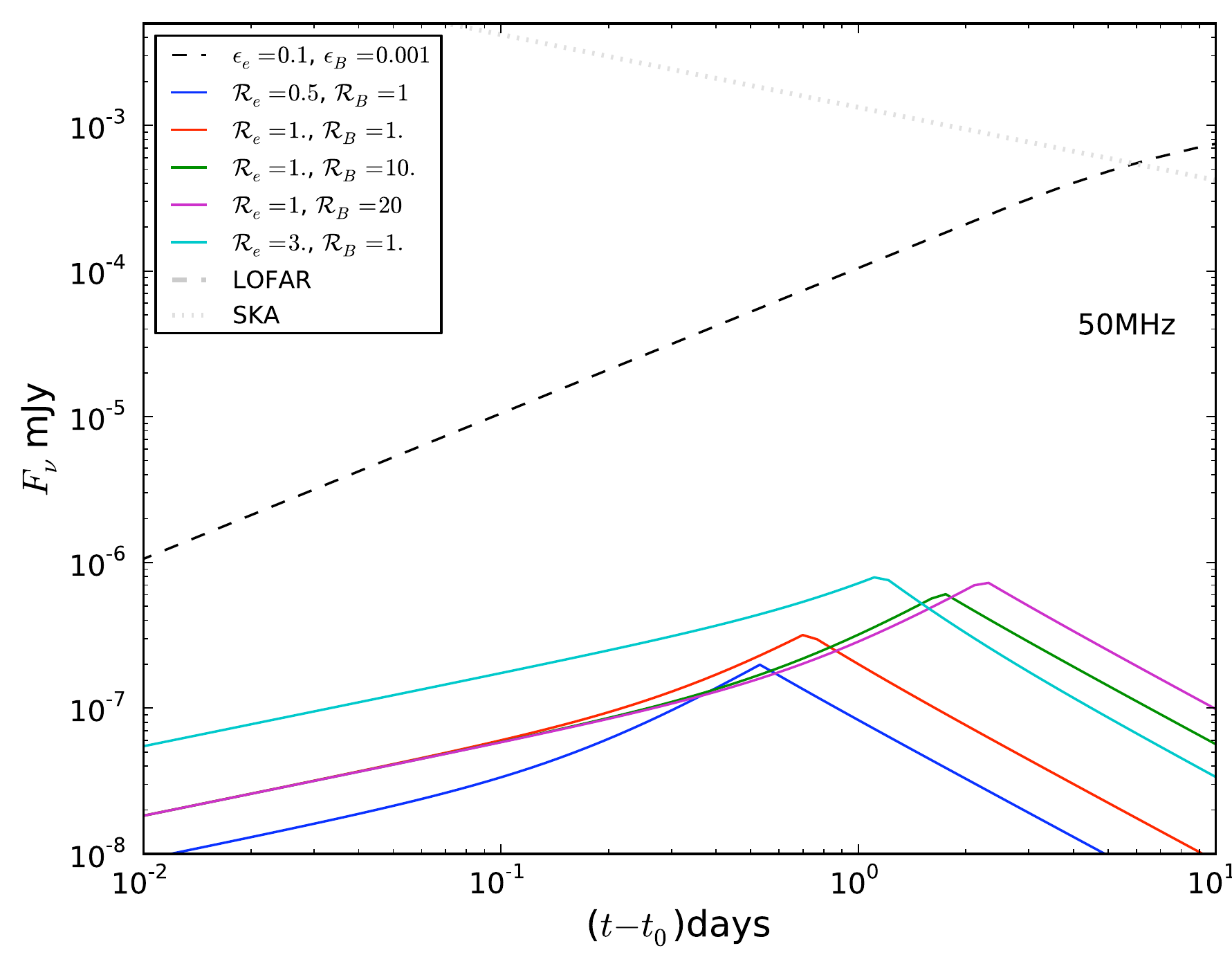}
\caption{The thick shell wind model. Physical parameters are $E_{iso,52} = 5.$, $\eta = 300$, $A_{\star} = 0.01$ and $T = 200$s. Again, the FS flux is nearly constant for GHz frequencies due to the spectral regime in which GHz frequency falls. In GHz frequencies, the RS peak corresponds to the fireball becoming optically thin, while for lower frequencies, the peak correspond to change in self-absorption ($\nuaRS = \nuaFS$).}
\end{center}
\end{figure*}
\begin{figure*}
\begin{center}
\includegraphics[scale=0.42]{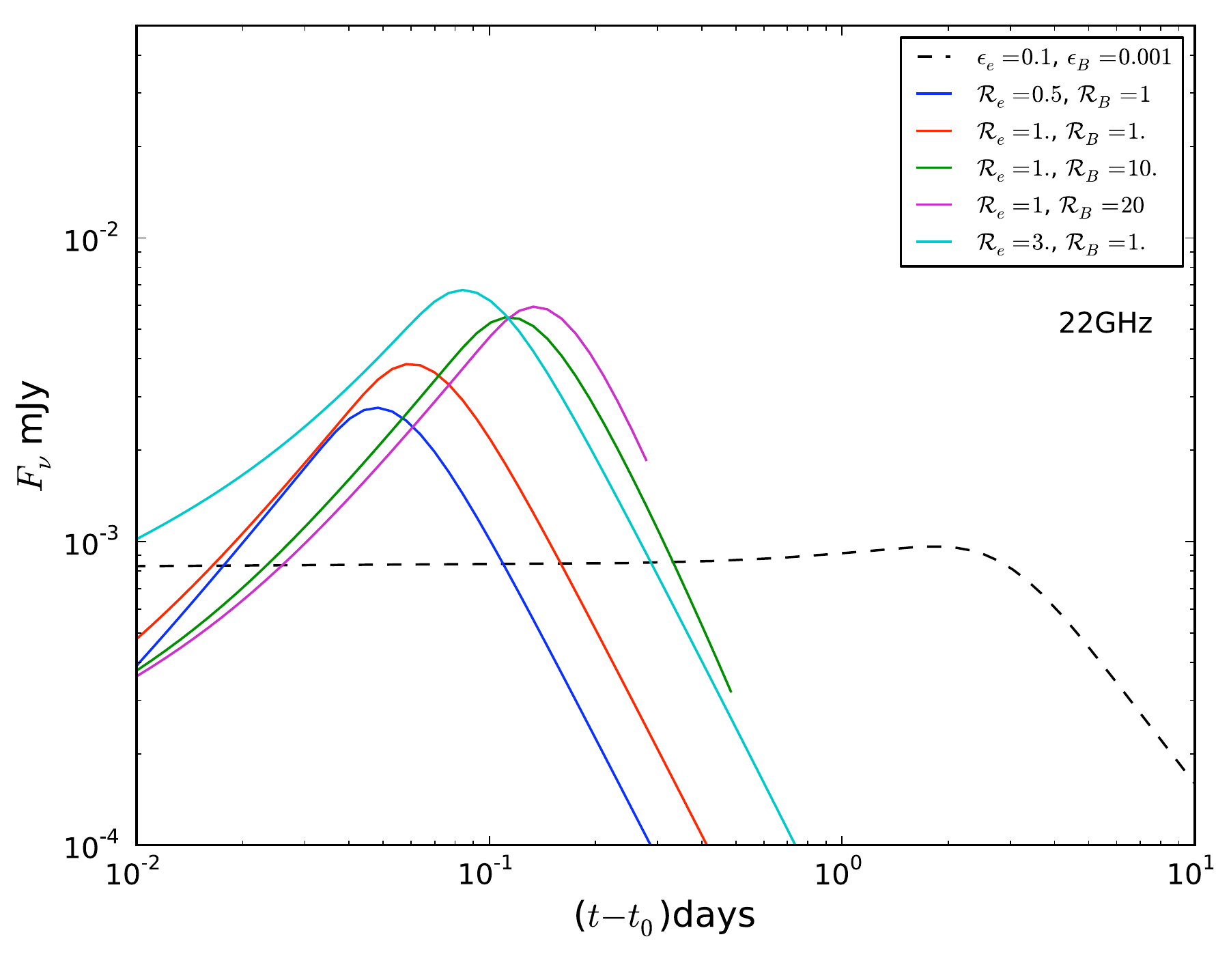}
\includegraphics[scale=0.42]{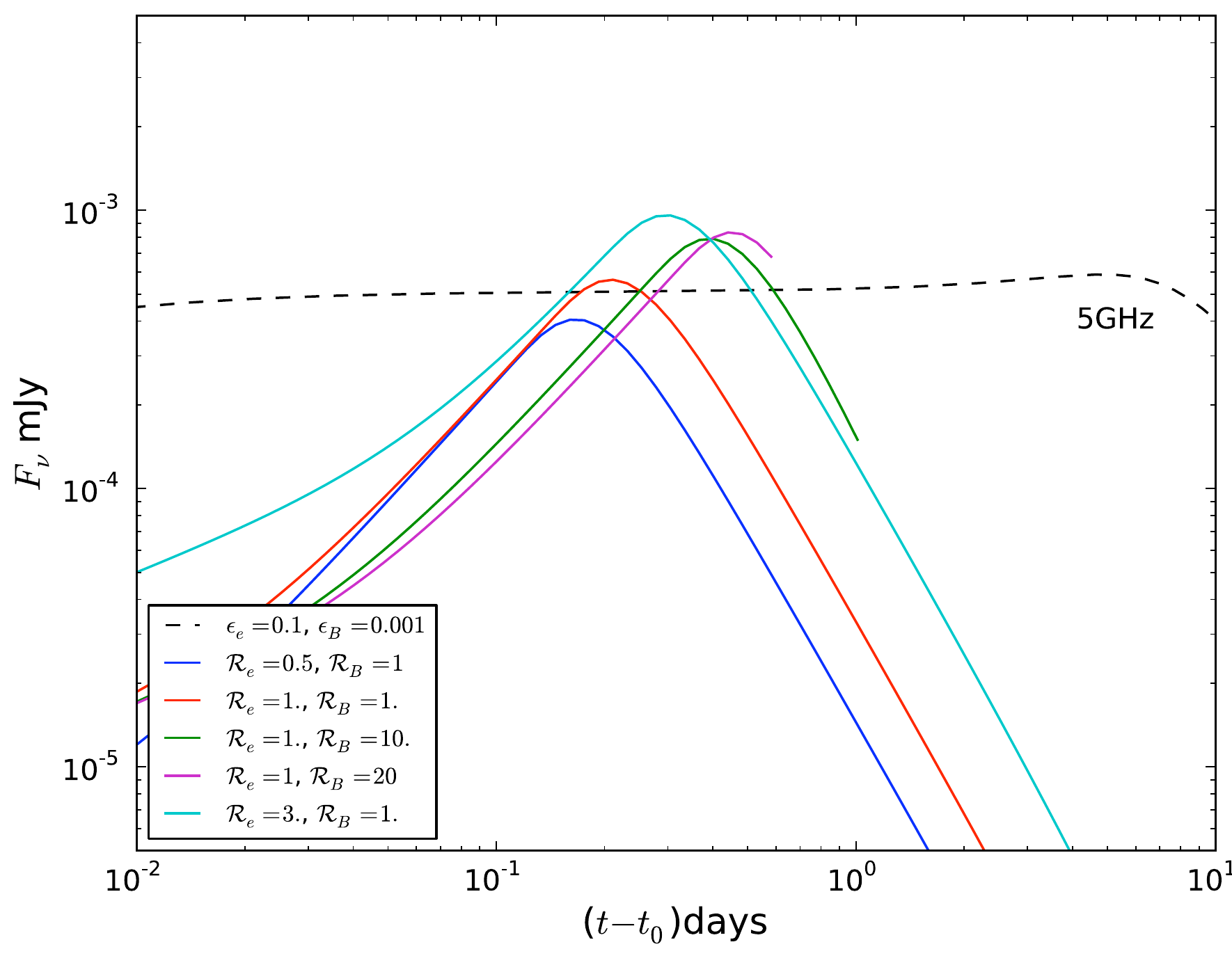}
\includegraphics[scale=0.30]{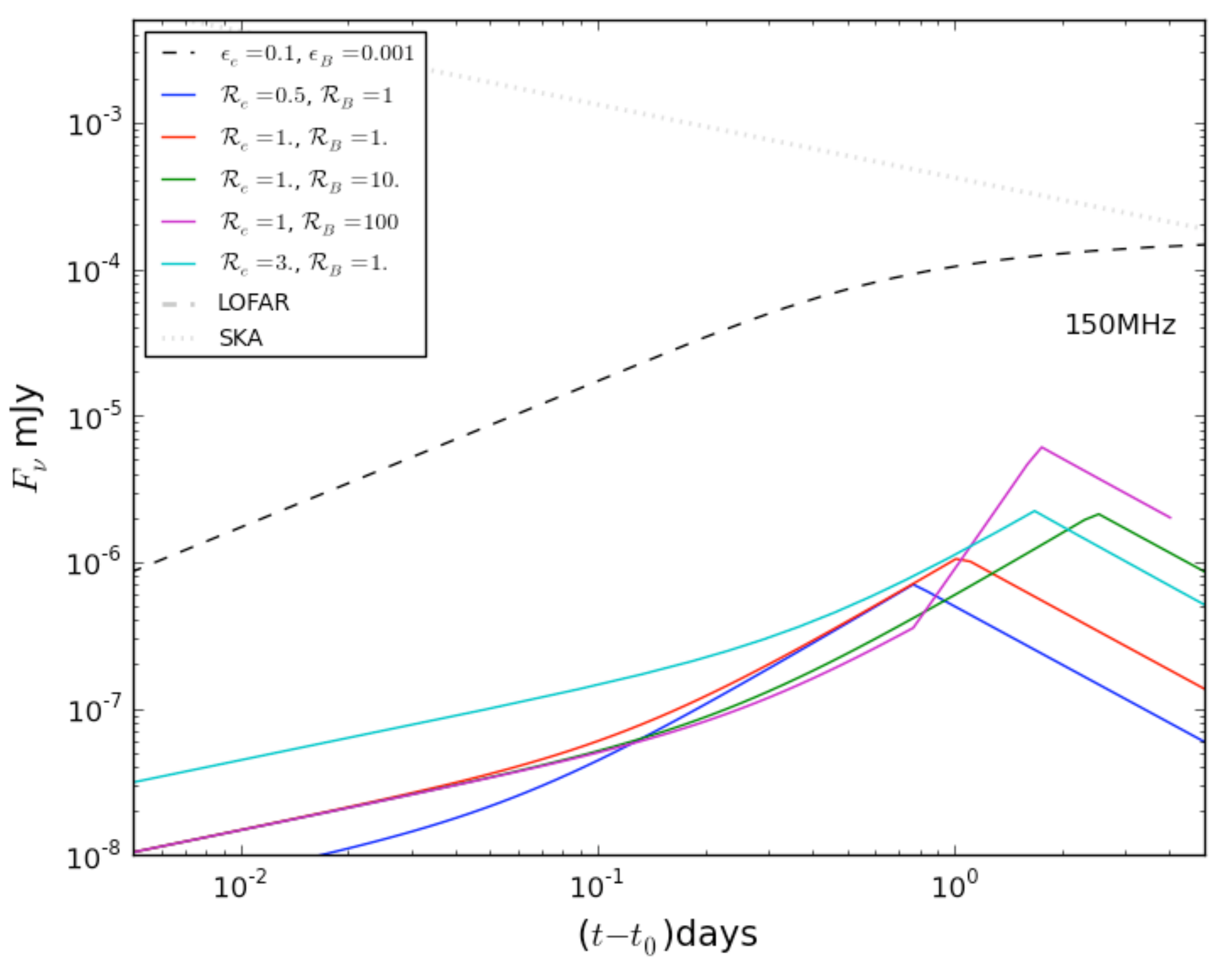}
\includegraphics[scale=0.42]{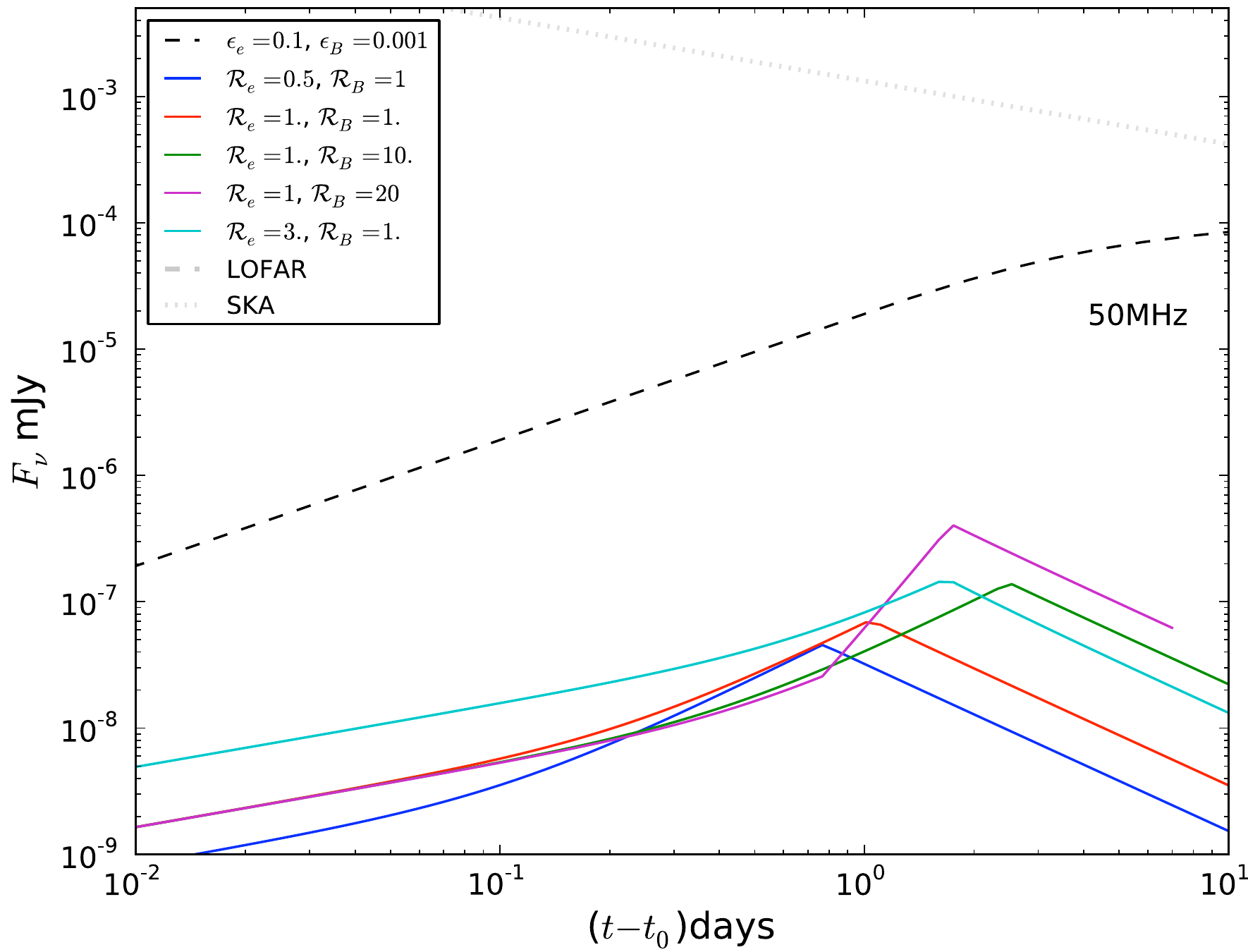}
\caption{Same as previous figure, but for $z=5$.}
\end{center}
\end{figure*}
\begin{figure*}
\begin{center}
\includegraphics[scale=0.23]{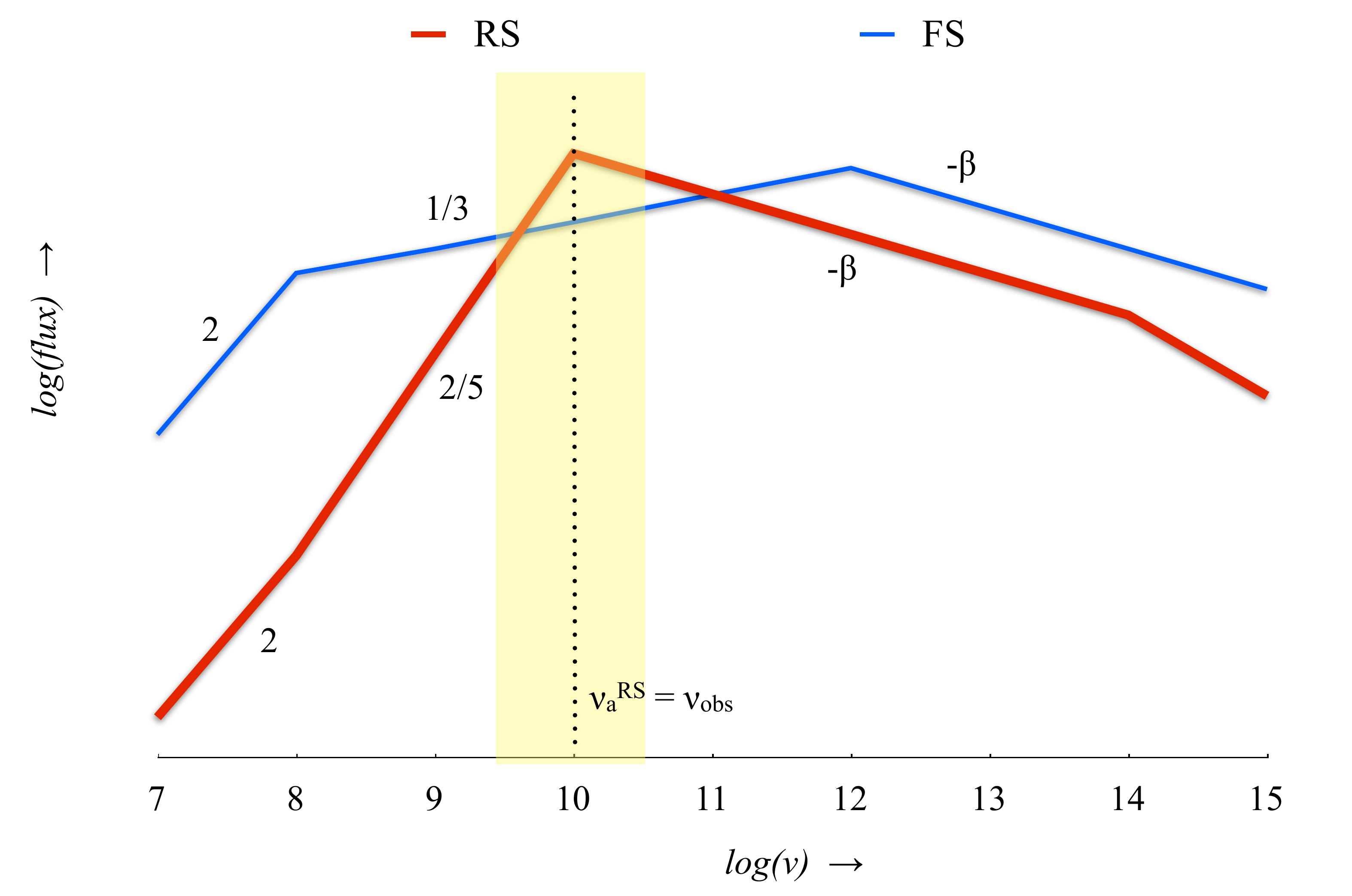}
\includegraphics[scale=0.23]{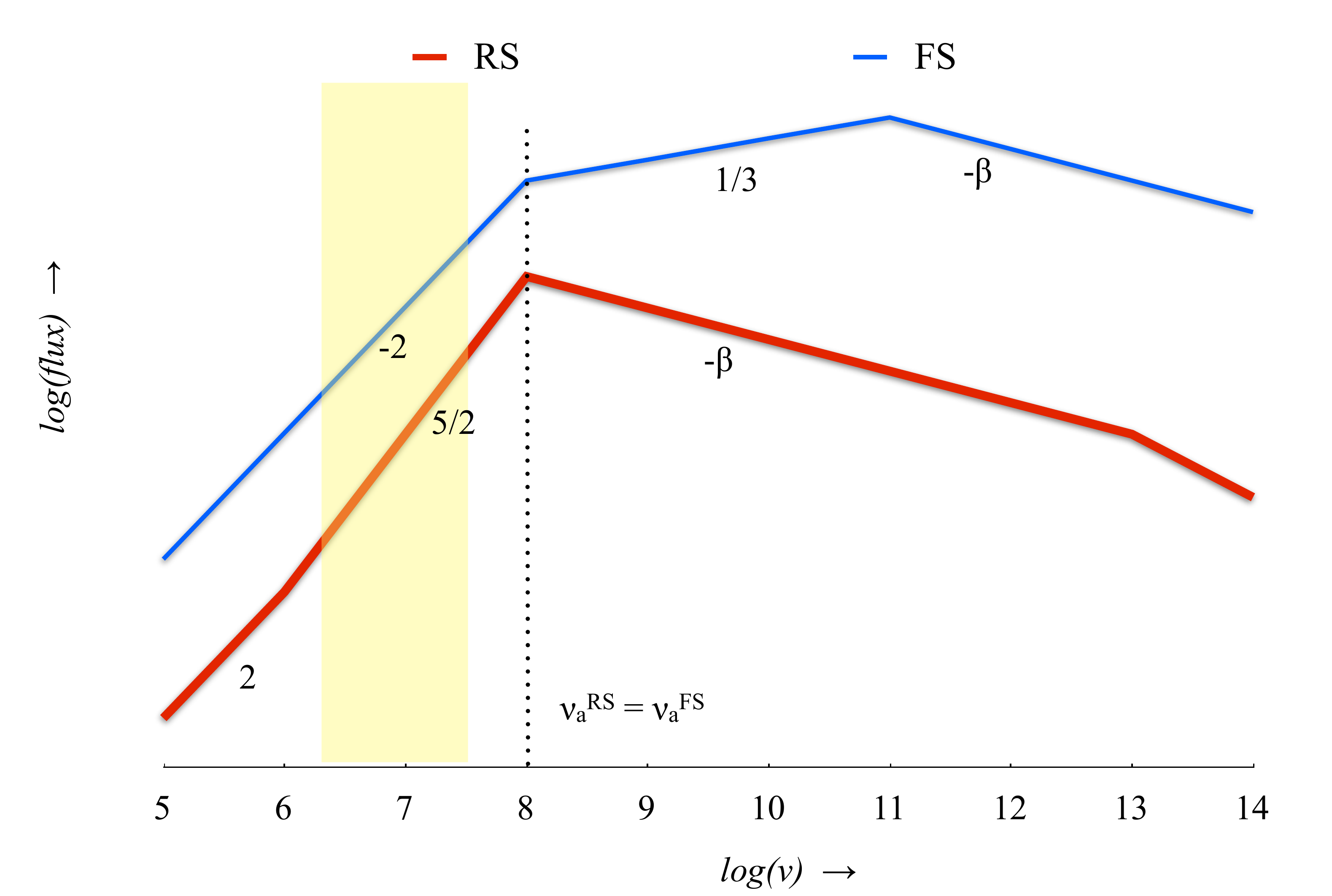}
\caption{ Sketch of FS (grey)and RS (red) spectrum for typical physical parameters at \textit{(left:)} $t_a$, the peak of GHz frequency RS lightcurve and \textit{(right:)}  $t_{eq}$, the peak of MHz frequency RS lightcurve. Yellow-highlighted region is the range observing frequencies. We can see that $t_{eq}$ is much later than $t_a$. Hence between the two epoches, the FS $f_m$ either remains constant (ISM) or falls off as $t^{-1}$ (wind). The RS $f_m$ falls far steeper and thereby reduces the RS flux at $t_{eq}$ compared to $t_a$. We have not considered the flux suppression due to a jet break.}
\end{center}
\end{figure*}

\begin{figure}
\begin{center}
\includegraphics[scale=0.42]{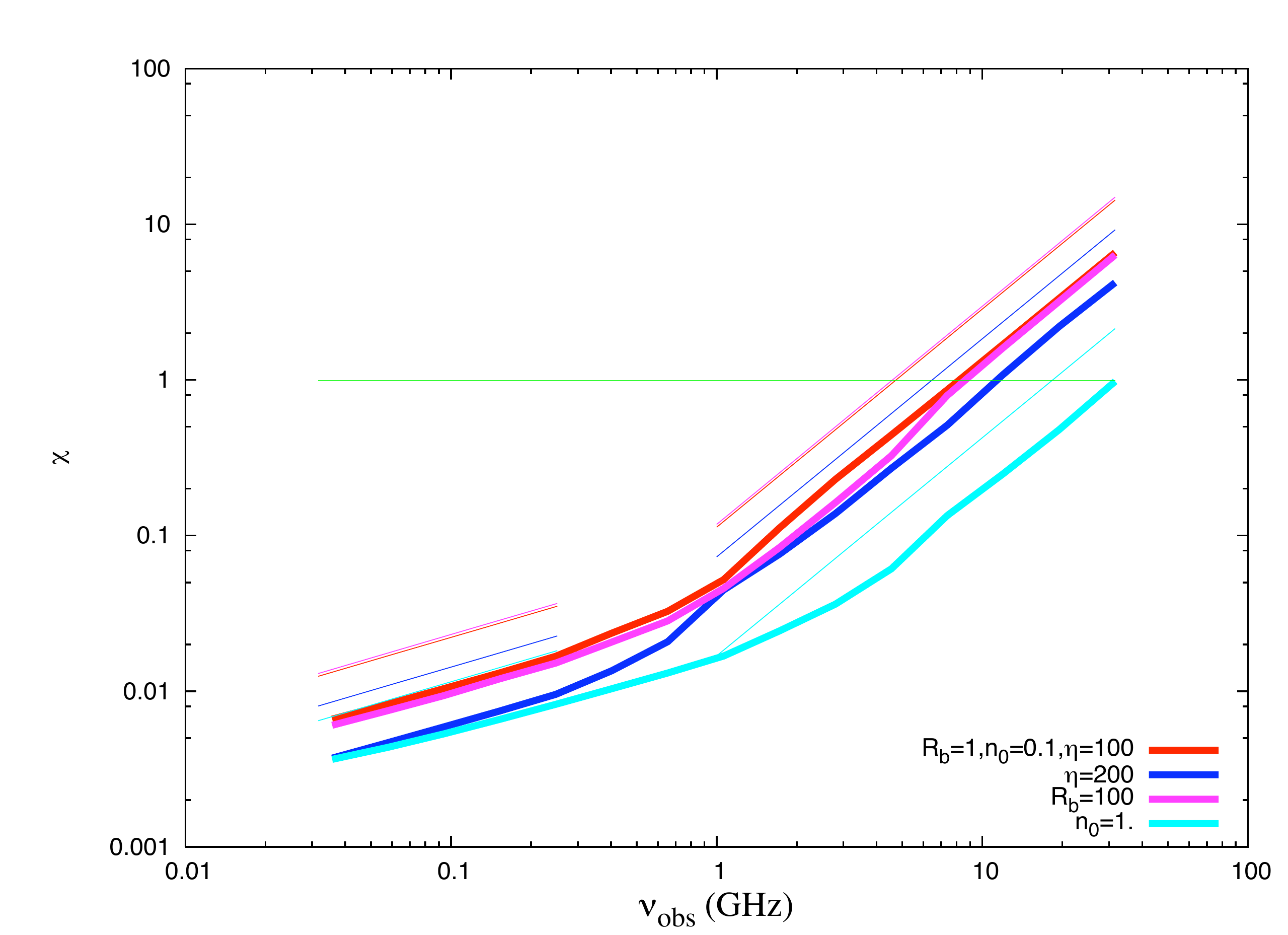}
\caption{$\chi$ vs $\nu_{\rm obs}$ for different sets of parameters for the thin ISM case. Thick lines are numerical results and thin lines are analytical approximations using formulae in section 5.2. The difference between the two owes to the analytical approximations in evaluating the thermal energy density and number density in shocked ejecta (RS), smoothing of synchrotron spectrum in numerical calculations and possibility of a jet break at $\tp$ in low frequencies. Red ($\mathcal{R}_{\rm B} = 1, \eta = 100, n_0 = 0.1$), blue ($\mathcal{R}_{\rm B} = 1, \eta = 200, n_0 = 0.1$), magenta ($\mathcal{R}_{\rm B} = 100, \eta = 100, n_0 = 0.1$), cyan ($\mathcal{R}_{\rm B} = 1, \eta = 200, n_0 = 1.0$). Other parameters remain the same ($E_{52} = 5., \mathcal{R}_e =1, \epsilon_e = 0.1, \epsilon_{\rm B} = 1.e-3$, $\theta_j = 5.^{\circ}$ and $z = 1$.) The horizontal line marks the detectability of reverse shock ($\chi =1$.)}
\end{center}
\end{figure}

\begin{figure}
\begin{center}
\includegraphics[scale=0.42]{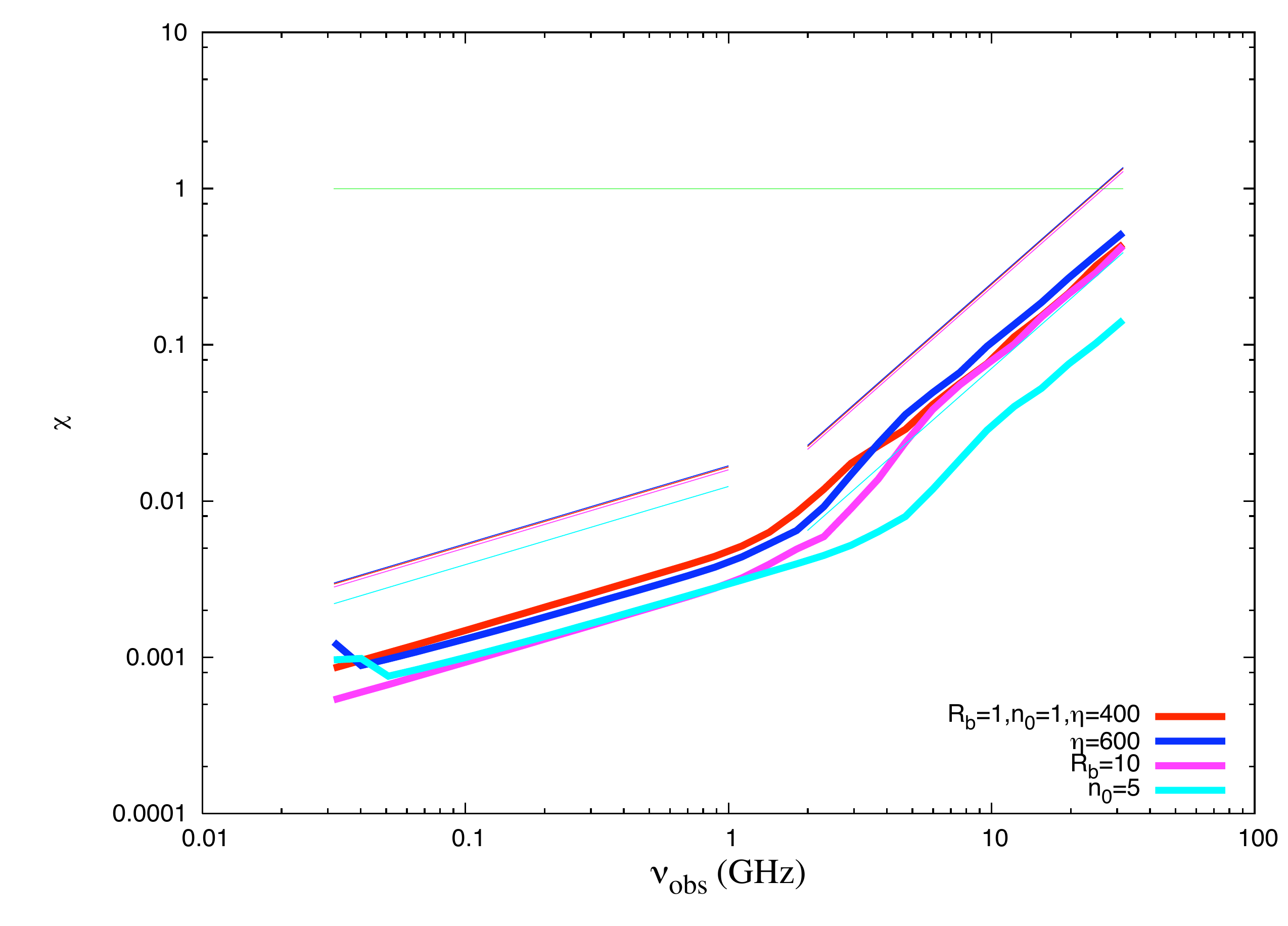}
\caption{Same as Fig.10, but for the thick ISM case. Red ($\mathcal{R}_{\rm B} = 1, \eta = 400, n_0 = 1.$), blue ($\mathcal{R}_{\rm B} = 1, \eta = 600, n_0 = 1.$), magenta ($\mathcal{R}_{\rm B} = 10, \eta = 400, n_0 = 1.$), cyan ($\mathcal{R}_{\rm B} = 1, \eta = 400, n_0 = 5.$). Other parameters remain the same ($E_{52} = 1., \mathcal{R}_e =1, \epsilon_e = 0.1, \epsilon_{\rm B} = 1.e-3, T_{90} = 80$ and $z=1$).}
\end{center}
\end{figure}

\begin{figure}
\begin{center}
\includegraphics[scale=0.42]{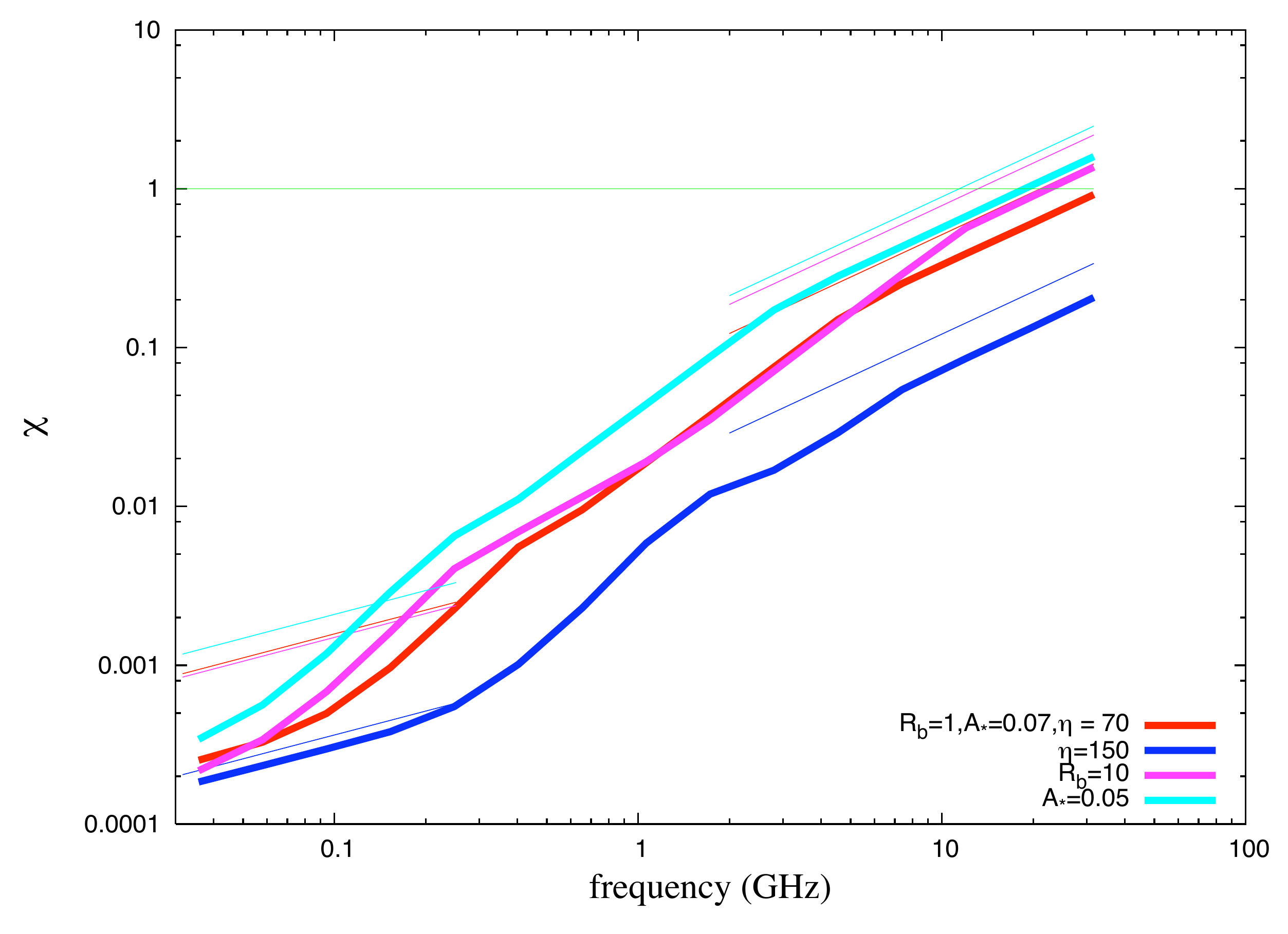}
\caption{Same as Fig.10, but for the thin wind case. Red ($\mathcal{R}_{\rm B} = 1, \eta = 70, A_{\star} = 0.07$), blue ($\mathcal{R}_{\rm B} = 1, \eta = 150, A_{\star} = 1.$), magenta ($\mathcal{R}_{\rm B} = 10, \eta = 70, A_{\star} = 0.07$), cyan ($\mathcal{R}_{\rm B} = 1, \eta = 70, A_{\star} = 0.05$). Other parameters remain the same ($E_{52} = 5., \mathcal{R}_e =1, \epsilon_e = 0.1, \epsilon_{\rm B} = 8.e-5$ and $z=1$).}
\end{center}
\end{figure}

\begin{figure}
\begin{center}
\includegraphics[scale=0.42]{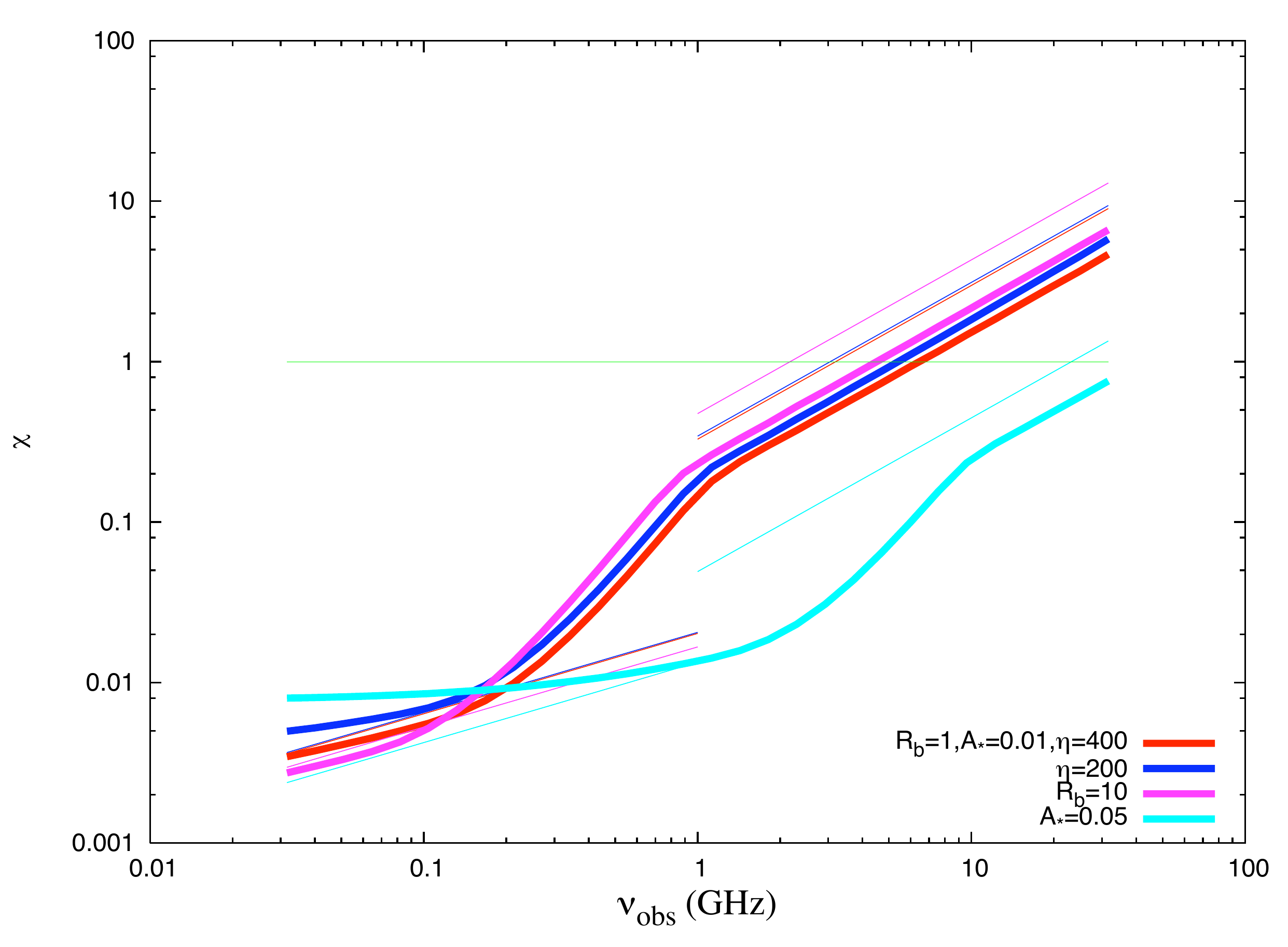}
\caption{Same as previous figure, but for the thick Wind case. Red ($\mathcal{R}_{\rm B} = 1, \eta = 400, A_{\star} = 0.01$), blue ($\mathcal{R}_{\rm B} = 1, \eta = 700, A_{\star} = 1.$), magenta ($\mathcal{R}_{\rm B} = 10, \eta = 400, A_{\star} = 1.$), magenta ($\mathcal{R}_{\rm B} = 1, \eta = 400, A_{\star} = 0.05$). Other parameters remain the same ($E_{52} = 1., \mathcal{R}_e =1, \epsilon_e = 0.1, \epsilon_{\rm B} = 1.e-3, T_{90} = 80$ and $z=1$). For the low frequency part, the spectral regime of numerical estimate is different from that assumed in analytical expressions.}
\end{center}
\end{figure}

\end{document}